\documentclass[useAMS,usenatbib]{mn2e}
\usepackage{graphicx}  
\usepackage{float}

\title[Early-stage star forming cloud cores in EGOs as traced by organic species]
      {Early-stage star forming cloud cores in {\it GLIMPSE} Extended Green Objects (EGOs) as traced by organic species}

\author[J. X. Ge, J. H. He, X. Chen, and S. Takahashi]
{ J. X. Ge$^{1,2}$ \thanks{E-mail: gejixing@ynao.ac.cn },
  J. H. He$^{1}$ \thanks{E-mail: jinhuahe@ynao.ac.cn },
  X. Chen$^{3,4}$  and S. Takahashi$^{5,6,7}$  \\
$^{1}$ Key Laboratory for the Structure and Evolution of Celestial Objects,
  Yunnan observatories, Chinese Academy of Sciences, \\
  P.O. Box 110, Kunming, 650011, Yunnan Province, PR China\\
$^{2}$ Chinese Academy of Sciences University, PR China\\
$^{3}$ Key Laboratory for Research in Galaxies and Cosmology,
  Shanghai Astronomical Observatory, Chinese Academy of Sciences, \\
  80 Nandan Road, Shanghai 200030, PR China \\
$^{4}$ Key Laboratory of Radio Astronomy, Chinese Academy of Sciences, 
  West Beijing Road, Nanjing, Jiangsu 210008, PR China\\
$^{5}$ Joint ALMA Observatory, Alonso de C{\'o}rdova 3107, Vitacura, Santiago, Chile\\
$^{6}$ National Astronomical Observatory of Japan, 2-21-1 Osawa, Mitaka, Tokyo 181-8588, Japan \\
$^{7}$ Academia Sinica, Institute of Astronomy and Astrophysics,
  P.O. Box 23-141, Taipei 10617, Taiwan\\ 
}

\begin{document}


\pagerange{\pageref{firstpage}--\pageref{lastpage}} \pubyear{2014}

\maketitle

\label{firstpage}

\begin{abstract}
 In order to investigate the physical and chemical properties of massive star forming cores in early stages, we analyse the excitation and abundance of four organic species, CH$_3$OH, CH$_3$OCH$_3$, HCOOCH$_3$ and CH$_3$CH$_2$CN, toward 29 Extended Green Object (EGO) cloud cores that were observed by our previous single dish spectral line survey. The EGO cloud cores are found to have  similar methanol $J_3-J_2$ rotation temperatures of $\sim 44$\,K, a typical linear size of $\sim 0.036$\,pc, and a typical beam averaged methanol abundance of several $10^{-9}$ (the beam corrected value could reach several $10^{-7}$). The abundances of the latter three species, normalized by that of methanol, are found to be correlated also across a large variety of clouds such as EGO cloud cores, hot corinos, massive hot cores and Galactic Center clouds. The chemical properties of the EGO cloud cores lie between that of hot cores and hot corinos. However, the abundances and abundance ratios of the four species can not be satisfactorily explained by recent chemical models either among the EGO cloud cores or among the various types of cloud cores from literature.

\end{abstract}


\section{Introduction}

The formation processes of massive stars are still poorly understood mainly due to their complexity, short evolution time scale, and the lack of nearby samples. But it has been recognised that it is not a simple scaled-up version of low-mass star formation \citep{zinn2007}. The differences may have occurred during the cloud core formation in the early stage of star formation. Many recent observational \citep[see e.g., ][]{zhang2009,Pere2013} and theoretical \citep[see e.g., ][]{smith2009,kauf2010} efforts have been focused on this topic. 

The environments of forming massive stars largely depend on the formation and evolution of their natal giant molecular clouds (GMCs). Various studies have demonstrated that the GMCs may be transient objects that are constantly forming and dispersing, and only a small fraction of the cloud mass becomes gravitationally bound to make star forming cores \citep[see e.g., ][]{vazq2005,dobb2006,elme2007,ward2014}. The different stages of the core collapse and warm-up are best traced by their distinct chemical features. There have been some chemical simulation works upon the cloud core chemistry and evolution \citep[e.g.,][]{aika2001,rodg2001,rodg2003,garr2006,garr2008}. The gas and grain chemistry is usually driven by the evaporation of various key parent species such as CO, H$_2$CO, NH$_3$, H$_2$O, and CH$_3$OH at various gas temperatures. There have been chemical observations upon both early starless dense cores \citep[e.g.,][]{vand2005,crap2005,schn2007,sakai2010b,bour2012} and later hot molecular cloud cores harboring forming massive stars \citep[e.g.,][]{mook2007,biss2007,qin2010,sakai2013,xu2013}. However, observation of larger samples of massive star forming cores at various evolutionary stages are still needed.

The discovery of the Extended Green Objects or Green Fuzzies \citep{cyga2008,cham2009,chen2013a,chen2013b} has greatly extended the available sample of massive star forming clouds. These objects possibly possess outflows that are traced by both the infrared excess emission around $4.5\,\micron$ and radio masers \citep[see][]{cyga2009,cyga2012b,cyga2013,chen2009,chen2011}. They are generally at the early stages of massive star formation, which is supported by their infrared colors and the lack of 1.3 or 3.6\,cm radio continuum source \citep{cyga2011b}. Thermal molecular line emission has also been explored toward them and rich chemistry has been reported \citep{chen2010,cyga2011a,he2012,paron2012,yu2013,kend2013}.

We have performed a systematic single dish millimeter spectral line survey toward all the 89 EGOs that have their declinations higher than $-38\degr$ in the catalogue of \citet{cyga2008} and detected the organic species CH$_3$OH in 38 of them \citep[][Paper I]{he2012}. Some more complex species are also detected in some objects. Paper I has demonstrated that most of these EGOs are associated with dense gas and many of them also possess active outflow shocks. The EGO clouds are also found to show uniform correlations among various spectral line luminosities but varying correlations between line widths. In this paper, we investigate the excitation status and abundances of the organic species in these EGO cloud cores and discuss their physical and chemical properties, so as to reveal the massive star forming environments therein.

This paper is organized as follows: We describe the observational data and data analysis approach in Sect.~\ref{obseveddata} and \ref{data_analysis}, respectively; The resulting parameters are given in Sect.~\ref{results}; Then, Sect.~\ref{discuss} concentrates on discussing the physical and chemical properties of the EGO clouds and comparing them with other types of interstellar clouds; Finally, a summary is given in Sect.~\ref{summary}.

\section[]{Observational data}
\label{obseveddata}
The data used in this work have been published in Paper\,I. The observations were performed using a 10\,m single dish telescope, the Heinrich Hertz Submillimeter Telescope (SMT), with a beam of about $29\arcsec$ around the wavelength of $1.1$\,mm. In this survey, all molecular lines of one and the same object were obtained by simultaneous exposures (in the upper and lower sidebands), which exempts the usual 20 per cent of nominal flux calibration error from the relative errors of the different lines.

In total, four organic species, CH$_3$OH, CH$_3$OCH$_3$, HCOOCH$_3$, and CH$_3$CH$_2$CN, were detected or tentatively detected in 38, 9, 7 and 5 objects, respectively, among the surveyed 89 EGOs. To foot our analysis upon good quality data, we abandon 9 objects whose methanol lines are so weak in the data that no more than two lines are detected. Thus, the total number of objects in our methanol abundance analysis is 29.

Due to the high frequency density of multiple transitions of these complex molecules, line blending frequently becomes an issue. Contributions of the blended lines were usually not differentiated in paper\,I. In this work, we judge the major contributors to the blended lines by comparing their strengths with adjacent transitions that do not suffer from blending, using non-blending objects as spectral templates. If only one of the blended transitions is the major contributor, the blended line is assigned to this transition and the other is taken as a non-detection. With this rule, CH$_3$OCH$_3$ 17$_{5,13}$-17$_{4,14}$EE in G35.20-0.74, and HCOOCH$_3$ 21$_{3,18}$-20$_{3,17}$\,A line in G12.91-0.26 and G34.41+0.24 have been taken as non-detections. If, however, both blended transitions could be non-negligible contributors, we arbitrarily divide the line strength into two halves with one half for each. This is the case for the blending between CH$_3$OCH$_3$ 16$_{5,11}$-16$_{4,12}$ EE and HCOOCH$_3$ 21$_{8,14}$-20$_{8,13}$\,E, and CH$_3$OCH$_3$ 17$_{5,13}$-17$_{4,14}$EE and HCOOCH$_3$ 21$_{5,17}$-21$_{5,16}$\,E in objects G12.91-0.26, G24.33+0.14, and G34.41+0.24, and between CH$_3$CH$_2$CN 29$_{5,26}$-28$_{5,25}$ and CH$_3$CH$_2$CN 29$_{4,26}$-28$_{4,25}$ in G12.68-0.18.

\section{Data analysis method}
\label{data_analysis}

\subsection{The population diagram analysis}
\label{pdanalysis}

Trial analysis has shown that the methanol lines in our objects are not always optically thin. Therefore, we adopt the population diagram (PD) analysis method from \citet{gold1999}, instead of the traditional rotation diagram (RD) analysis which is only suitable for optically thin lines. The only assumptions made in the PD method is the local thermodynamic equilibrium (LTE) in a homogeneous medium. PD also offers the capability to determine optical depth and filling factor of the transitions. We use the following PD formula from \citet{gold1999} to relate total column density $N_{\rm tot}$ (or optical depth $\tau$), rotational temperature $T_{\rm rot}$ and filling factor $f$ to each other
\begin{equation}
 \label{PDA}
 \ln(\frac{N_{u}^{thin}}{g_{u}})=\ln(\frac{N_{tot}}{Q})-
     \ln(\frac{1}{f})-\ln (C_{\tau}) - \frac{E_{u}}{kT_{rot}},
\end{equation}
where $g_u$ and $E_u$ are the statistic weight and the energy of the upper level, $Q$ is the partition function, $k$ is Boltzmann constant. The correction factor for line center opacity $\tau$ is 
\begin{equation}
C_{\tau}=\frac{\tau}{1-e^{-\tau}}.
\end{equation}
The upper level column density derived from observed integrated line intensities by assuming optically thin lines can be computed by 
\begin{equation}
   N_{u}^{thin}=\frac{8\pi k \nu^{2} W}{hc^{3}A_{ul}}=\gamma_{u}\cdot W.
\end{equation}
Here, $\nu$ is the lab frequency,
$W$ is the integrated line intensity (over velocity),
$h$ is the Plank constant,
$c$ is the speed of light. 
We use cgs units for all quantities. The Einstein coefficient 
$A_{ul}=1.16395\times 10^{-20} \nu^{3}S\mu^{2}/g_{u}$, with line strength and electric dipole moment $S\mu^{2}$ taken from CDMS
database\footnote{CDMS: http://www.astro.uni-koeln.de/cdms}. 
For the line center optical depth $\tau$, we use the formula from \citet{wang2010}:
\begin{equation}
\label{tau}
 \tau =\frac{c^{3}}{8 \pi \nu^{3}} \frac{A_{ul}}{\Delta V} \frac{g_{u} N_{tot}} {Q}
 e^{-\frac{ E_{u} }{kT_{rot}}}(e^{\frac{h\nu}{kT_{rot}}}-1),
\end{equation}
where $\Delta V$ is the full width at half maximum(FWHM) in cm\,s$^{-1}$ of the observed lines.

We fit the multiple transitions of a molecule by reduced $\chi^2$ minimization. For any given combination of $N_{\rm tot}$, $T_{\rm rot}$ and $f$, an expected integrated line area  $W_i$ can be predicted for $i$'th line using the above equations, which are then compared with the observed ones to define 

\begin{equation}
\chi^2=\frac{1}{n-p}\sum_{i=1}^{n} (\frac{ W_{i,{\rm mod}}-W_{i,{\rm obs}} }{ \sigma_{i,{\rm obs}} })^2
\end{equation}
where $\sigma_{i,{\rm obs}}$ is the uncertainty of the observed line area, $n$ is the number of observed lines and $p$ is the number of free parameters. The line width  $\Delta V$ in Eq.~(\ref{tau}) is taken as the average FWHM of all involved observed lines. The best solution is searched for in a large enough parameter space of $N_{\rm tot}=10^{13}-10^{19}$ cm$^{-2}$, $T_{\rm rot}=10-300$ K, and $f=0.00001-1.0$. We then flexibly zoom into local regions of the parameter space to fine tune the solution domain to find the minimum value $\chi_{\rm min}^2$ corresponding to the best solution. It is expected that $\chi_{\rm min}^{2}<<1$ when the data are over-fitted while $\chi_{\rm min}^{2}>>1$ when the data are not adequately represented by the fitting model.

We assume that the homogeneous LTE cloud model is adequate for the EGO clouds and all the spectral line measurements follow normal distributions, so that we can figure out various solution domains corresponding to typical confidence levels of 68.3 per cent ($1\sigma$), 90 per cent ($1.6\sigma$) and 99 per cent ($2.6\sigma$). The contour level of a solution domain can be defined by an increment $\Delta \chi^2$ as 
\begin{equation}
\chi^2=\chi_{\rm min}^2+\frac{\Delta \chi^2}{n-p}.
\end{equation}
According to \citet{lamp1976}, the increment $\Delta \chi^2$ equals  
to 3.5 ($1\sigma$), 6.25 ($1.6\sigma$) and 11.3 ($2.6\sigma$) respectively when all three parameters are fitted ($p=3$) and 
2.3 ($1\sigma$), 4.61 ($1.6\sigma$), and 9.21 ($2.6\sigma$) respectively when one of the parameters is fixed ($p=2$). 

We stress that the parameter uncertainties determined above are very crude, because other sources of uncertainties such as the limitation of the PD method (particularly when $\chi_{\rm min}^{2}>>1$ or $\chi_{\rm min}^{2}<<1$), parameter degeneracy (see detailed discussions below) have not been accounted for in a self-consistent way.

\subsection{parameter degeneracy}
\label{para_dengeneracy}

According to \citet{gold1999} and \citet{wang2010}, the PD method works the best when the observed lines involve both optically thin and thick lines. When the involved lines are all thin or thick, parameter degeneracy occurs. In our sample, nine objects have their methanol lines belonging to the best case. Thus, all three parameters can be uniquely determined and we call these objects group-I in this work. An example contour plot is given in Fig.~\ref{fig1} in Appendix~\ref{examplefig} for G35.20-0.74 in this group.

In the case where all lines are totally optically thick, the opacity correction factor becomes $C_\tau\approx \tau$. Then Eq.~(\ref{PDA}) becomes 
\begin{equation}
 N_{u}^{thin}\approx\frac{g_{u}N_{tot}f}{Q\tau}e^{-\frac{E_{u}}{kT_{rot}}}
=\frac{8\pi k\nu^2}{hc^3}\frac{\Delta V}{A_{ul}} T_{\rm b}f
\end{equation}
where we have applied Eq.~(\ref{tau}) and 
\begin{equation}
 \label{Tb}
 T_{\rm b} = \frac{h\nu}{k}\frac{1}{e^{\frac{h\nu}{kT_{\rm rot}}}-1}
\end{equation}
is the brightness temperature. It is clear that the information of the column density is totally lost and the brightness temperature is completely degenerate with the filling factor. Three of our objects have their methanol lines falling into this case and we call these objects group-II. However, we recognize that the methanol lines of our objects in this group are only slightly optically thick. Thus, the brightness temperature and filling factor are only partially degenerate with each other. It is still possible to obtain unique solutions for them by taking their partial degeneracy as a part of their uncertainties. Concerning the column density, it is still possible to obtain a lower limit for it from the PD contour maps, because it must be high enough to guarantee optically thick condition for all involved lines. An example contour plot is given in Fig.~\ref{fig2} in Appendix~\ref{examplefig} for G23.01-0.41 in this group.

In the case where all lines are optically thin, the opacity correction factor $C_\tau$ becomes unity. Then Eq.~(\ref{PDA}) becomes 
\begin{equation}
 \label{RDA}
 \ln(\frac{N_{u}^{thin}}{g_{u}})=\ln(\frac{N_{tot}f}{Q})-\frac{E_{u}}{kT_{rot}},
\end{equation}
which is exactly the formula of rotation diagram (RD) analysis and $N_{tot}f$ is the beam averaged column density. Therefore, the column density and the filling factor are completely degenerate. Seventeen of our objects have their methanol lines belonging to this case and we call these objects group-III. We fix the filling factor to $f=1.0$ to directly compute beam averaged quantities. We also estimate the filling factor of these objects by adopting an average methanol cloud core linear size from the group-I objects (see details in Sect.~\ref{massdensity}), so that we can get rough estimates of their column densities and line opacities. An example contour plot is given in Fig.~\ref{fig3} in Appendix~\ref{examplefig} for G10.34-0.14 in this group.

\subsection{The faint lines case}
\label{baddata}

The lines of the other three organic species, CH$_3$OCH$_3$, HCOOCH$_3$ and CH$_3$CH$_2$CN, are usually much weaker than that of methanol. They have fewer detected lines and their population diagrams look more incomplete. In this case, we fix the rotation temperature to some empirical value and the filling factor to unity so that we only need to determine the beam averaged column density. We stress that we still apply the PD method to these faint lines, because it yields a better estimation of the column density than the rotation diagram method by handling the opacity effect in a self-consistent way. 

The typical evaporation temperature of methanol, the parent species of the three organics, from grain surfaces in cloud warm-up chemical models \citep{garr2008} is about 90\,K. However, as we shall discuss later in this work, the methanol rotation temperature  that is computed as the average among all groups-I, II and III objects as 44\,K can be close to the thermal temperature of the gas,  although we can not preclude the possibility of underestimating the gas temperature due to the non-thermal excitation of methanol. Thus, we adopt this temperature for the PD analysis of the three organic molecules, but also test the computation at two extreme $T_{\rm rot}$ of 19 and 83\,K. The latter tests show that the derived column densities of CH$_3$OCH$_3$ only change by 15 to 70 per cent, while that of HCOOCH$_3$ can change by factors of $2\sim 3$ and that of CH$_3$CH$_2$CN by factors of $3\sim 6$. 

\subsection{Outlier lines}
\label{outliers}

\begin{table}
 \centering
 \begin{minipage}{80mm}  
  \caption{List of a few unexpectedly too strong CH$_3$OH lines that are excluded from our population diagram analysis.}
  \label{tab1}
  \begin{tabular}{@{}lll@{}}
  \hline
   object    &  transition(Lower Sideband) & $\nu$ (MHz) \\
  \hline
   G10.34-0.14 & CH$_3$OH 9$_{3,7}$ -- 9$_{2,8}$ +-  & 252089.14 \\
   G12.20-0.03 & CH$_3$OH 8$_{3,5}$ -- 8$_{2,6}$ -+  & 251515.65 \\
   G12.42+0.50 & CH$_3$OH 8$_{3,5}$ -- 8$_{2,6}$ -+  & 251517.47 \\
   G18.89-0.47 & CH$_3$OH 9$_{3,7}$ -- 9$_{2,8}$ +-  & 252092.37 \\
   G19.01-0.03 & CH$_3$OH 8$_{3,5}$ -- 8$_{2,6}$ -+  & 251518.08 \\
   G19.01-0.03 & CH$_3$OH 8$_{3,6}$ -- 8$_{2,7}$ +-  & 251983.84 \\
   G19.01-0.03 & CH$_3$OH 10$_{3,8}$ -- 10$_{2,9}$ +- & 252252.58 \\
   G19.01-0.03 & CH$_3$OH 11$_{3,9}$ -- 11$_{2,10}$ +- & 252486.30 \\
   G22.04+0.22 & CH$_3$OH 3$_{3,0}$ -- 3$_{2,1}$ -+ & 251905.81 \\
   G22.04+0.22 & CH$_3$OH 9$_{3,7}$ -- 9$_{2,8}$ +- & 252090.99 \\
   G34.26+0.15 & CH$_3$OH 10$_{3,8}$ -- 10$_{2,9}$  +-  & 252252.64 \\
   G35.03+0.35 & CH$_3$OH 9$_{3,7}$ --  9$_{2,8}$ +- & 252088.95 \\
   G35.13-0.74 & CH$_3$OH 10$_{3,8}$ -- 10$_{2,9}$ +- & 252251.47 \\
   G35.79-0.17 & CH$_3$OH 9$_{3,7}$ -- 9$_{2,8}$ +- & 252090.65 \\
   G35.79-0.17 & CH$_3$OH 10$_{3,8}$ -- 10$_{2,9}$ +- & 252251.78 \\
   G35.79-0.17 & CH$_3$OH 11$_{3,9}$ -- 11$_{2,10}$ +- & 252484.04 \\
   G54.11-0.08 & CH$_3$OH 8$_{3,5}$ -- 8$_{2,6}$ -+ & 251518.10 \\
\hline
\end{tabular} \\ [1mm]
\end{minipage}
\end{table}

After some tests, we find that some individual CH$_{3}$OH lines are significantly stronger than expected from an LTE population diagram. These outlier lines are the ones with the highest level energies and usually with poor S/N ratios. Thus, we have excluded them from our PD analysis (see the list in Table~\ref{tab1}). However, we do not exclude the probability that they would be caused by non-LTE effects or multiple temperature components \citep[see, e.g.,][]{numm2000}.

\section{Results}
\label{results}
The resulting parameters of all four organic species, CH$_3$OH, CH$_3$OCH$_3$,
HCOOCH$_3$ and CH$_3$CH$_2$CN, from the population diagram analysis are presented in Tables~\ref{tab_ch3oh} and \ref{tab_other3}. They are described in detail below.

\subsection{Methanol (CH$_3$OH)}
\label{ch3oh}

\begin{table*}
\centering
\begin{minipage}{160mm}  
\caption{Parameters of CH$_3$OH from population diagram analysis.}
\label{tab_ch3oh}
\begin{tabular}{@{}lcccccccc@{}}
\hline
\hline
Object &  $N_{\rm l}$ &$\chi^2_{\rm min}$&  $T_{\rm rot}$ & $f$ & $N_{\rm tot}$& $N_{\rm tot}\times f$& $X^a$ & $\tau^b$\\
       &       & & & $\times 0.01$  &$\times 10^{17}$&$\times 10^{14}$ &$\times 10^{-9}$ & \\
       &       & & (K) &   &(cm$^{-2}$)& (cm$^{-2}$)      &                     &  \\
 (1)   & (2)          &  (3)           &  (4)&  (5)         & (6)  & (7)   &(8)   &(9)  \\
\hline
\multicolumn{9}{|c|}{Group I ($\tau\sim 1$) } \\
G11.92-0.61 & 15 & 2.8 & 59$_{-33}^{+29}$ & 0.17$_{-0.07}^{+0.23}$ & 3.49$_{-1.97}^{+39.00}$ & 6.09$_{-4.22}^{+68.50}$ & 1.65$_{-1.14}^{+18.51}$ & 0.95-2.82\\
G12.91-0.26 & 15 & 2.8 & 57$_{-5}^{+5}$ & 0.62$_{-0.06}^{+0.06}$ & 2.58$_{-0.56}^{+0.63}$ & 16.03$_{-3.78}^{+4.20}$ & 1.72$_{-0.41}^{+0.45}$ & 0.79-2.52\\
G14.33-0.64 & 15 & 3.2 & 37$_{-2}^{+3}$ & 1.69$_{-0.27}^{+0.56}$ & 0.61$_{-0.23}^{+0.26}$ & 10.26$_{-4.21}^{+5.58}$ & 3.80$_{-1.56}^{+2.07}$ & 0.14-1.24\\
G16.59-0.05 & 15 & 2.5 & 48$_{-10}^{+9}$ & 0.39$_{-0.06}^{+0.11}$ & 3.24$_{-1.23}^{+2.89}$ & 12.66$_{-5.18}^{+11.84}$ & 5.75$_{-2.36}^{+5.38}$ & 0.92-4.31\\
G34.41+0.24 & 15 & 4.3 & 58$_{-6}^{+6}$ & 0.66$_{-0.08}^{+0.08}$ & 3.57$_{-0.69}^{+0.86}$ & 23.57$_{-5.38}^{+6.33}$ & 9.82$_{-2.24}^{+2.64}$ & 1.12-3.37\\
G35.13-0.74 & 12 & 1.2 & 27$_{-11}^{+10}$ & 0.59$_{-0.18}^{+0.48}$ & 1.61$_{-1.16}^{+9.24}$ & 9.52$_{-7.44}^{+55.17}$ & 7.32$_{-5.73}^{+42.43}$ & 0.20-2.95\\
G35.20-0.74 & 15 & 4.4 & 58$_{-6}^{+5}$ & 0.60$_{-0.03}^{+0.05}$ & 2.53$_{-0.52}^{+0.52}$ & 15.26$_{-3.23}^{+3.40}$ & 13.87$_{-2.94}^{+3.09}$ & 0.79-2.42\\
G35.79-0.17 & 12 & 2.9 & 28$_{-12}^{+13}$ & 0.33$_{-0.13}^{+0.52}$ & 1.85$_{-1.56}^{+13.50}$ & 6.03$_{-5.62}^{+45.04}$ & 3.17$_{-2.96}^{+23.70}$* & 0.10-2.81\\
G40.28-0.22 & 15 & 1.2 & 39$_{-6}^{+5}$ & 0.58$_{-0.08}^{+0.10}$ & 1.67$_{-0.58}^{+0.99}$ & 9.75$_{-3.63}^{+6.00}$ & 6.50$_{-2.42}^{+4.00}$ & 0.35-2.71\\
\multicolumn{9}{|c|}{Group II ($\tau \ga 1$) } \\
G12.68-0.18 & 15 & 3.2 & 83$_{-63}^{+201}$ & 0.24$_{-0.19}^{+0.82}$ & $>$0.83  & $>$7.77  & $>$4.09* & $>$0.99\\
G23.01-0.41 & 15 & 2.5 & 46$_{-23}^{+149}$ & 0.32$_{-0.25}^{+0.46}$ & $>$7.16  & $>$44.59  & $>$8.26 & $>$2.56\\
G24.33+0.14 & 15 & 1.9 & 81$_{-21}^{+213}$ & 0.17$_{-0.13}^{+0.07}$ & $>$2.35  & $>$8.17  & $>$4.30* & $>$1.61\\
\multicolumn{9}{|c|}{Group III ($\tau \la 1$) } \\
G10.34-0.14 & 11 & 2.6 & 30$_{-3}^{+4}$   & 1.62$^\dagger$ & 0.16$^\dagger$ & 2.58$_{-0.31}^{+0.38}$ & 1.43$_{-0.17}^{+0.21}$ & 0.15$^\dagger$\\
G12.20-0.03 &  5 & 0.3 & 38$_{-9}^{+20}$  & 0.32$^\dagger$ & 0.59$^\dagger$ & 1.90$_{-0.42}^{+0.58}$ & 1.06$_{-0.23}^{+0.32}$ & 0.86$^\dagger$\\
G12.42+0.50 &  8 & 1.0 & 25$_{-4}^{+5}$   & 1.22$^\dagger$ & 0.14$^\dagger$ & 1.67$_{-0.34}^{+0.54}$ & 1.52$_{-0.30}^{+0.49}$ & 0.32$^\dagger$\\
G14.63-0.58 &  4 & 3.0 & 49$_{-18}^{+67}$ & 1.42$^\dagger$ & 0.08$^\dagger$ & 1.12$_{-0.18}^{+0.50}$ & 1.40$_{-0.23}^{+0.63}$ & 0.09$^\dagger$\\
G18.89-0.47 &  4 & 0.3 & 35$_{-11}^{+34}$ & 0.34$^\dagger$ & 0.27$^\dagger$ & 0.90$_{-0.22}^{+0.40}$ & 0.53$_{-0.13}^{+0.24}$ & 0.55$^\dagger$\\
G19.01-0.03 &  8 & 0.4 & 30$_{-6}^{+9}$   & 0.38$^\dagger$ & 0.25$^\dagger$ & 0.97$_{-0.16}^{+0.25}$ & 0.39$_{-0.06}^{+0.10}$ & 0.34$^\dagger$\\
G19.36-0.03 & 12 & 1.2 & 59$_{-12}^{+22}$ & 1.14$^\dagger$ & 0.10$^\dagger$ & 1.17$_{-0.11}^{+0.13}$ & 0.17$_{-0.02}^{+0.02}$ & 0.12$^\dagger$\\
G19.88-0.53 & 15 & 2.0 & 50$_{-3}^{+4}$   & 0.59$^\dagger$ & 0.52$^\dagger$ & 3.07$_{-0.12}^{+0.12}$ & 6.14$_{-0.23}^{+0.24}$ & 0.60$^\dagger$\\
G22.04+0.22 & 11 & 0.5 & 71$_{-17}^{+33}$ & 0.51$^\dagger$ & 0.27$^\dagger$ & 1.41$_{-0.14}^{+0.33}$ & 0.45$_{-0.04}^{+0.11}$ & 0.16$^\dagger$\\
G23.96-0.11 &  7 & 0.2 & 33$_{-10}^{+24}$ & 0.34$^\dagger$ & 0.29$^\dagger$ & 0.99$_{-0.19}^{+0.38}$ & 0.52$_{-0.10}^{+0.20}$* & 0.57$^\dagger$\\
G24.00-0.10 &  6 & 0.8 & 47$_{-21}^{+201}$& 0.35$^\dagger$ & 0.23$^\dagger$ & 0.82$_{-0.14}^{+2.71}$ & 0.22$_{-0.04}^{+0.71}$ & 0.29$^\dagger$\\
G25.38-0.15 &  7 & 1.4 & 24$_{-4}^{+6}$   & 0.24$^\dagger$ & 0.71$^\dagger$ & 1.69$_{-0.38}^{+0.67}$ & 0.37$_{-0.08}^{+0.15}$ & 1.66$^\dagger$\\
G34.26+0.15 & 11 & 4.2 & 30$_{-2}^{+2}$   & 0.49$^\dagger$ & 0.74$^\dagger$ & 3.59$_{-0.27}^{+0.31}$ & 1.38$_{-0.10}^{+0.12}$ & 1.25$^\dagger$\\
G35.03+0.35 & 11 & 1.0 & 40$_{-6}^{+8}$   & 0.57$^\dagger$ & 0.36$^\dagger$ & 2.05$_{-0.22}^{+0.25}$ & 1.58$_{-0.17}^{+0.19}$ & 0.34$^\dagger$\\
G45.47+0.05 &  7 & 1.5 & 20$_{-5}^{+8}$   & 0.26$^\dagger$ & 1.00$^\dagger$ & 2.54$_{-0.97}^{+2.62}$ & 2.31$_{-0.88}^{+2.38}$ & 1.96$^\dagger$\\
G54.11-0.08 &  3 & 0.4 & 19$_{-5}^{+11}$  & 0.44$^\dagger$ & 0.35$^\dagger$ & 1.57$_{-0.85}^{+2.58}$ & 1.96$_{-1.07}^{+3.22}$ & 0.42$^\dagger$\\
G59.79+0.63 &  5 & 0.6 & 30$_{-10}^{+24}$ & 0.36$^\dagger$ & 0.40$^\dagger$ & 1.42$_{-0.33}^{+0.89}$ & 1.09$_{-0.25}^{+0.69}$ & 0.56$^\dagger$\\
\hline
\hline
\end{tabular} \\ [1mm]
Note: The objects are grouped according to their line opacity case as described in Sect.~\ref{para_dengeneracy}.
The columns are
(1) Object name;
(2) Number of detected lines;
(3) Minimum value of $\chi^2$;
(4) Rotational temperature $T_{\rm rot}$;
(5) Filling factor $f$;
(6) Column density $N_{\rm tot}$ corrected for filling factor $f$;
(7) Beam-averaged column density $N_{\rm tot}\times f$;
(8) Beam-averaged relative abundance $X$;
(9) Optical depth $\tau$ (ranges for group I, lower limits for group II and median values for group III objects).\\
{\em a}: $X$ is computed using $N_{\rm tot}\times f$ in column (7) and $N$(H$_2$) from Chen et al (2010).\\
{\em b}: The $\tau$ range of group-I objects is for individual lines, the $\tau$ lower limits of group-II objects are determined from population diagram analysis, while the median $\tau$ of group-III objects is obtained with assumed linear object size (see details in Sect.~\ref{para_dengeneracy}).\\
{\em *}: $N$(H$_2$) is unavailable in \citet{chen2010} and assumed to the average value of 1.9$\times 10^{23}$ cm$^{-2}$ of EGOs clouds from that work.\\
{\em $\dagger$}: The column densities and the median methanol line opacities are estimated using the filling factor calculated from assumed average cloud core linear size of 0.036 pc (See the discussion of the optically thin case in Sect.~\ref{para_dengeneracy})
\end{minipage}
\end{table*}

The population diagram analysis is performed to 29 EGOs that have no less than three clearly detected CH$_3$OH lines. The line frequencies, lower level energies, the values of $S\mu^{2}$
and partition function of CH$_3$OH are taken from CDMS database. The molecular hydrogen column densities $N({\rm H}_2$) of the EGO clouds that are needed for computing abundances are estimated by \citet{chen2010} using the C$^{18}$O 1-0 line. The results are listed in Table~\ref{tab_ch3oh}.

In general, the $\chi_{\rm min}^2$ values are not far from unity, suggesting that our PD fits are basically good. However, there is a clear trend that opaque objects in groups-I and II tend to have $\chi_{\rm min}^2>1$ while optically thin objects in group-III tend to have $\chi_{\rm min}^2<1$. There could be slight non-LTE effects in the opaque objects while optically thin cases are over-fitted to some degree.

The methanol rotation temperatures of these objects are in the range of 19-83\,K, with average values of 46 and 38\,K for the optically thinner objects in groups-I and III, and 71\,K for the optically thicker objects in group-II. These temperatures roughly agree with previously observed massive SFRs \citep[e.g., 30-200\,K were found toward 13 massive SFRs by][]{tak2000}.

The filling factors of the objects in groups-I and II are on the order of $10^{-3}$, which means that the cloud cores that possess organic contents are much smaller than our telescope beam ($29\arcsec$). The beam averaged column densities ($N_{tot}\times f$) vary from a few $10^{13}$\,cm$^{-2}$ up to higher than a few $10^{15}$\,cm$^{-2}$, with higher values in optically thicker objects in groups-I and II. True column densities or their lower limits can be estimated for all 29 EGOs to be about $10^{16}\sim 10^{17}$\,cm$^{-2}$ using the known or estimated filling factors, which are three orders of magnitude higher than the beam averaged ones. These values roughly agree with the CH$_3$OH column densities in dense cloud cores \citep[e.g., AFGL\,2591, G24.78, G75.78, NGC\,6334\,IRS1, NGC\,7538\,IRS1, W\,3(H$_2$O) and W\,33A, ][]{biss2007}.

The beam averaged methanol abundances (column 8) are in the range of 10$^{-10}-10^{-8}$, with higher values in optically thicker objects in groups-I and II. The average beam averaged abundance in all 29 objects is about $3.2\times 10^{-9}$.

The optical depths (column 9) of individual methanol lines are around unity for group-I objects, which agrees with the fact that they can be best analysed with the population diagram method without the problem of parameter degeneracy \citep{gold1999,wang2010}. The line optical depth lower limits in group-II and the median depths in group-III are also consistently roughly higher and lower than unity, respectively. There are few exceptional group-III objects showing median line opacities larger than unity, slightly contradicting with our object grouping based on the characteristics of their population digrams, perhaps due to their small numbers of data points and non-LTE effects.

\subsection{Dimethyl ether (CH$_3$OCH$_3$)}
\label{ch3och3}
\begin{table}
 \centering
 \begin{minipage}{80mm} 
  \caption{Parameters of CH$_3$OCH$_3$, HCOOCH$_3$ and CH$_3$CH$_2$CN determined with a fixed rotation temperature of $T_{\rm rot}=44$\,K and a fixed filling factor of $f=1.0$.}
  \label{tab_other3}
  \begin{tabular}{l@{ }c@{  }cll@{}c}
  \hline
  \hline
Object & $N_{\rm l}$ & $\chi^2_{\rm min}$  & $N_{\rm tot}\times f$  & X $^a$           & Grp \\
       &             &                     & $\times 10^{14}$       & $\times 10^{-9}$ &    \\
(1)    &  (2)        &  (3)                &   (4)                  & (5)              & (6)\\
  \hline
CH$_3$OCH$_3$:  & & & &\\
G12.68-0.18 & 2 &  1.44 &  2.06$_{- 0.45 }^{+ 0.44}$ &   1.08$_{- 0.24 }^{+ 0.23}$* & II \\
G12.91-0.26 & 6 & 14.45 &  4.91$_{- 0.37 }^{+ 0.37}$ &   0.53$_{- 0.04 }^{+ 0.04}$ & I \\
G14.33-0.64 & 2 &  0.25 &  3.25$_{- 0.43 }^{+ 0.43}$ &   1.20$_{- 0.16 }^{+ 0.16}$ & I \\
G16.59-0.05 & 2 &  0.12 &  1.74$_{- 0.33 }^{+ 0.34}$ &   0.79$_{- 0.15 }^{+ 0.15}$ & I \\
G24.33+0.14 & 2 &  0.43 &  5.04$_{- 0.97 }^{+ 0.96}$ &   2.65$_{- 0.51 }^{+ 0.50}$* & II \\
G34.41+0.24 & 5 & 11.32 &  4.54$_{- 0.39 }^{+ 0.39}$ &   1.89$_{- 0.16 }^{+ 0.16}$ & I \\
\hline
HCOOCH$_3$: & & & &\\
G12.91-0.26 & 4 & 11.50 & 11.40$_{- 1.16 }^{+ 1.13}$ &   1.23$_{- 0.12 }^{+ 0.12}$ & I \\
G24.33+0.14 & 2 &  7.39 &  4.61$_{- 1.02 }^{+ 1.03}$ &   2.43$_{- 0.54 }^{+ 0.54}$* & II \\
G34.41+0.24 & 5 & 39.96 & 20.30$_{- 1.31 }^{+ 1.31}$ &   8.46$_{- 0.55 }^{+ 0.55}$ & I \\
G35.20-0.74 & 3 &  4.92 & 19.90$_{- 2.35 }^{+ 2.38}$ &  18.09$_{- 2.14 }^{+ 2.16}$ & I \\
\hline
CH$_3$CH$_2$CN:  & & & &\\
G12.68-0.18 & 2 &  3.95 &  0.69$_{- 0.15 }^{+ 0.15}$ &  0.36$_{- 0.08 }^{+ 0.08}$* & II \\
G12.91-0.26 & 3 &  4.17 &  1.48$_{- 0.21 }^{+ 0.20}$ &  0.16$_{- 0.02 }^{+ 0.02}$ & I \\
G23.01-0.41 & 2 &  0.23 &  2.02$_{- 0.39 }^{+ 0.38}$ &  0.37$_{- 0.07 }^{+ 0.07}$ & II \\
G24.33+0.14 & 2 &  0.51 &  0.77$_{- 0.16 }^{+ 0.16}$ &  0.40$_{- 0.08 }^{+ 0.08}$* & II \\
G34.41+0.24 & 5 &  2.11 &  1.61$_{- 0.16 }^{+ 0.16}$ &  0.67$_{- 0.07 }^{+ 0.07}$ & I \\
\hline
\hline
\end{tabular} \\[1mm]
Note: The columns are
(1) Object name;
(2) Number of observed lines;
(3) Minimum value of $\chi^2$;
(4) Beam-averaged column density $N_{\rm tot}\times f$ (cm$^{-2}$);
(5) Beam-averaged relative abundance;
(6) Group ID of CH$_3$OH line opacity cases in Table~\ref{tab_ch3oh}.\\
{\em a}: The molecular hydrogen column densities $N$(H$_2$) from \citet{chen2010} is used in computing the abundances.\\
{\em *}: $N$(H$_2$) is unavailable in \citet{chen2010} and assumed to the average value of 1.9$\times 10^{23}$ cm$^{-2}$ of EGOs clouds from that work.\\
\end{minipage}

\end{table}

 The CH$_{3}$OCH$_{3}$ lines are detected in nine EGOs. Although the lines are usually not very strong and there is no good enough level energy coverage for a robust PD analysis, we still apply the PD method by fixing $T_{\rm rot}=44$\,K and $f=1.0$, as described in Sect.~\ref{baddata}. At lease two detected lines are required for fitting one free parameter. Thus, only six of the nine EGOs can have their CH$_{3}$OCH$_{3}$ lines analysed. The line frequencies $\nu$, lower level energies $E_{\rm l}$, $S\mu^2$ and partition function are taken from CDMS database. The partition function takes into account all four symmetry substates AA, EE, AE, and EA and the low-lying V$_{\rm t}$=0,1,2 vibrational states. The results are shown in the first section of Table~\ref{tab_other3}.

The $\chi_{\rm min}^2$ values vary in a much larger range than for methanol lines. Over-fitting occurs usually when there are only two lines. Very large $\chi_{\rm min}^2$ values occur when there are more lines available, which is due to large irregular scatter of data points on the population diagrams (we do not show the plots). We have tested and excluded our treatment of the few cases line blending as the cause of the large data scatter. One possible reason is that non-LTE effects could be strong. Thus, the parameters estimated for dimethyl ether are far less reliable than that of methanol.

 The average (beam averaged) column density and abundance of dimethyl ether among the six EGO clouds are $3.6\times 10^{14}$ cm$^{-2}$ and $1.4\times 10^{-9}$, respectively, which agree with the abundances of $\sim 10^{-9}$ in some northern IRDCs \citep{vasy2014}. The dimethyl ether lines are detected only in those EGO clouds with unity or higher opacity CH$_{3}$OH lines (objects in groups-I and II in Table~\ref{tab_ch3oh}) and the abundances of CH$_{3}$OCH$_{3}$ are lower than that of CH$_{3}$OH.

\subsection{Methyl formate (HCOOCH$_3$)}
\label{hcooch3}

The HCOOCH$_{3}$ lines are detected in seven EGOs. Similarly, we apply PD analysis by fixing $T_{\rm rot}=44$\,K and $f=1.0$. Only four of the seven EGOs have at least two HCOOCH$_{3}$ lines detected, which is required by the PD analysis with one free parameter. The line frequencies $\nu$, lower level energies $E_{\rm l}$, and $S\mu^2$ are taken from splatalogue database\footnote{Splatalogue database for astronomical spectroscopy, http://www.cv.nrao.edu/php/splat}, while the partition function is from JPL database\footnote{JPL line list is at\\ http://spec.jpl.nasa.gov/ftp/pub/catalog/catform.html}.
The partition function includes both the torsional substates A and E
and the low-lying V$_t$=0,1 vibrational states. The results are shown in the second section of Table~\ref{tab_other3}.

All $\chi_{\rm min}^2$ values are larger than unity. Very large  $\chi_{\rm min}^2$ values also occur, similarly due to irregular data scatter. Similar tests show that the large data scatter is not due to our treatment of few cases of line blending, but to other reasons such as non-LTE effects.

The average (beam averaged) methyl formate column density and abundance among the four EGO clouds are $1.4\times 10^{15}$ cm$^{-2}$ and $7.6\times 10^{-9}$, respectively. This abundance is higher than that of CH$_{3}$OCH$_{3}$ by a factor of more than five. Similarly, the methyl formate lines are detected only in objects with unity or higher opacity CH$_{3}$OH lines but the abundances of HCOOCH$_{3}$ are comparable to that of CH$_{3}$OH.

\subsection{Ethyl cyanide (CH$_3$CH$_2$CN)}
\label{ch3ch2cn}

Similarly, PD analysis is applied to five EGOs with CH$_3$CH$_2$CN detections by fixing $T_{\rm rot}=44$\,K and $f=1.0$. The line frequencies $\nu$, lower level energies $E_{\rm l}$, and $S\mu^2$ are taken from the splatalogue database while the partition function are adopted from CDMS database. The results are shown in the third section of Table~\ref{tab_other3}.
 
All the $\chi_{\rm min}^2$ values are not far from unity, mainly due to the small numbers of detected lines. Perhaps, the non-LTE effects in ethyl cyanide are similar as in methanol, but weaker than in the other two species.

The average (beam averaged) ethyl cyanide column density and abundance among the five EGO clouds are $\sim 1.3\times 10^{14}$ cm$^{-2}$ and $3.9\times 10^{-10}$, respectively. Similarly, the ethyl cyanide lines are detected only in objects with unity or higher opacity CH$_{3}$OH lines.
The ethyl cyanide abundances are the lowest among the four considered organic molecules in the detected objects.

\subsection{Size, mass and density of EGO cloud cores}
\label{massdensity}
\begin{table}
 \centering
 \begin{minipage}{80mm}  
  \caption{Physical parameters of EGO cloud cores.}
  \label{tab_ldsmn}
  \begin{tabular}{@{}crlcrr@{}}
  \hline
Object      & $L^a$           & $D  $      &  $S$    &     $M$      & $n$  \\
            &$\times 10^{23}$ &            &         &              & $\times 10^{6}$\\
            & (erg\,s$^{-1}$) & (kpc)      &  (pc)   &(M$_{\odot}$) & (cm$^{-3}$)    \\
 (1)        &   (2)           & (3)        &  (4)    &    (5)       &      (6)       \\
\hline
\multicolumn{6}{|c|}{Group I}      \\
G11.92-0.61 &  4.4 & 3.72 & 0.022 & 9.03 & 28.25 \\
G12.91-0.26 & 12.1 & 3.67$^b$ & 0.041 & 23.14 & 11.21 \\
G14.33-0.64 &  4.3 & 2.47 & 0.045 & 6.71 & 2.38 \\
G16.59-0.05 &  8.5 & 4.31 & 0.038 & 25.20 & 15.12 \\
G34.41+0.24 & 13.7 & 3.59 & 0.041 & 32.56 & 15.39 \\
G35.13-0.74 &  1.9 & 2.34 & 0.025 & 5.59 & 11.25 \\
G35.20-0.74 &  4.7 & 2.30 & 0.025 & 8.65 & 17.81 \\
G35.79-0.17 &  3.9 & 3.81 & 0.031 & 9.38 & 10.70 \\
G40.28-0.22 & 11.5 & 4.97$^b$ & 0.053 & 25.80 & 5.53 \\
\multicolumn{6}{|c|}{Group II}       \\
G12.68-0.18 & 10.3 & 4.59$^b$ & 0.032 & $>$4.52 & $>$4.63 \\
G23.01-0.41 & 12.0 & 4.58 & 0.036 & $>$51.27 & $>$34.83 \\
G24.33+0.14 & 12.3 & 5.93$^b$ & 0.034 & $>$15.02 & $>$12.10 \\
\multicolumn{6}{|c|}{Group III}       \\
G10.34-0.14 &  1.2 & 1.99 & 0.036$^c$ & 1.09 & 0.79 \\
G12.20-0.03 &  4.8 & 4.47 & 0.036$^c$ & 4.07 & 2.93 \\
G12.42+0.50 &  0.9 & 2.30 & 0.036$^c$ & 0.95 & 0.68 \\
G14.63-0.58 &  0.9 & 2.13 & 0.036$^c$ & 0.54 & 0.39 \\
G18.89-0.47 &  2.5 & 4.38 & 0.036$^c$ & 1.85 & 1.33 \\
G19.01-0.03 &  2.2 & 4.11 & 0.036$^c$ & 1.75 & 1.26 \\
G19.36-0.03 &  1.0 & 2.38 & 0.036$^c$ & 0.71 & 0.51 \\
G19.88-0.53 &  4.3 & 3.31 & 0.036$^c$ & 3.60 & 2.59 \\
G22.04+0.22 &  2.8 & 3.54 & 0.036$^c$ & 1.89 & 1.36 \\
G23.96-0.11 &  2.5 & 4.34 & 0.036$^c$ & 1.99 & 1.43 \\
G24.00-0.10 &  2.4 & 4.28 & 0.036$^c$ & 1.61 & 1.16 \\
G25.38-0.15 &  4.4 & 5.21 & 0.036$^c$ & 4.92 & 3.54 \\
G34.26+0.15 &  5.0 & 3.64 & 0.036$^c$ & 5.10 & 3.67 \\
G35.03+0.35 &  3.2 & 3.35$^b$ & 0.036$^c$ & 2.47 & 1.77 \\
G45.47+0.05 &  7.1 & 5.02 & 0.036$^c$ & 6.86 & 4.94 \\
G54.11-0.08 &  1.8 & 3.81 & 0.036$^c$ & 2.44 & 1.76 \\
G59.79+0.63 &  3.1 & 4.23 & 0.036$^c$ & 2.72 & 1.96 \\
\hline
\end{tabular} \\[1mm]
Note: The parameter columns are
(1) Object name;
(2) CH$_3$OH line luminosity;
(3) Distance;
(4) Cloud core diameter;
(5) Mass;
(6) Density.\\
{\em a}: The CH$_3$OH luminosities and distances are from Paper\,I. \\
{\em b}: The near kinematic distance is adopted.\\
{\em c}: The linear cloud core sizes of the group-III objects are fixed to the average value of the group-I objects.\\
\end{minipage}
\end{table}

The distances needed for estimating the physical parameters of the objects are available from Paper\,I. The kinematic distances were computed from spectral lines and the near and far kinematic distance ambiguity had been solved for most but five EGOs involved in this work. For the five undetermined objects, we adopt the near kinematic distances in our analysis. 

The linear size of the molecular cloud cores traced by the CH$_3$OH lines in the EGO clouds can be roughly estimated using the filling factor from the PD analysis. Usually, the filling factor is composed of two factors $f=f_b f_c$, where $f_b$ is the filling factor of the whole CH$_3$OH emission region in the telescope beam and $f_c$ is the clump filling factor inside the emission region. We assume that the clumpiness of the cloud core is not too prominent so that the surface brightness distribution is more or less smooth in the whole emission region. Then, we have $f_c\approx 1$ and $f\approx f_b$. Because the filling factors of our EGOs turn out to be always very small ($f\sim 10^{-3}$), the beam filling factor can be related to the angular size $\theta_s$ of the cloud core and the telescope beam size $\theta_b$ via $f_b=\theta_s^2/\theta_b^2$. Then, the linear size of a cloud core can be computed as 
  \begin{equation}
   S=\theta_{s}D \approx \theta_{b} \sqrt{f} D,
  \end{equation}
where $D$ is the distance to the object.

The aforementioned beam averaged abundance estimation suffers from great uncertainties in the H$_2$ column densities. Thus, we explore an alternative way to reveal the physical properties of the clouds: to estimate their gas mass and density by assuming a representative methanol abundance relative to H$_2$. For our sample of EGO clouds, the representative beam averaged CH$_{3}$OH abundance is taken to be the S/N weighted average $7.55\times 10^{-9}$ among the nine group-I objects (the objects with the most reliable abundances) in Table~\ref{tab_ch3oh}. As we will discuss later in Sect.~\ref{discuss_methanol}, after the H$_2$ column density in a typical EGO cloud core is approximately estimated, the methanol abundance in the core would be about 40 times higher than the beam averaged one on average. Thus, we will use this small-scale average methanol abundance $\overline{X}\approx 3.02\times 10^{-7}$ in the estimation of core mass and gas density. Then, the mass of the cloud core can be roughly estimated as
  \begin{equation}
  \label{mass}
   M=\frac{N_{tot}}{\overline{X}} (\pi R^2)(\mu_{\rm H_2} m_p)
    =\frac{N_{tot}}{\overline{X}} \frac{\pi}{4}S^2  \mu_{\rm{H_2}}m_p,
  \end{equation}
where $R=S/2$ is the radius of the core, $m_{p}$ is the proton mass and $\mu_{\rm{H_2}}=2.8$ is molecular weight per hydrogen molecule \citep{kauf2008}. The gas density can be estimated by adopting a spherical homogeneous cloud model with radius $R$ as 
 \begin{equation}
  \label{density}
  n=\frac{M}{(4/3)\pi R^3 \mu_p m_p}= \frac{N_{tot}}{\overline{X}} \frac{1}{S} \frac{3\mu_{\rm H_2}}{2\mu_p} .
 \end{equation}
where $\mu_p=2.37$ is the mean molecular weight per free particle.

The group-III objects with optically thin methanol lines in Table~\ref{tab_ch3oh} can not have their filling factor uniquely determined due to parameter degeneracy in the population diagram method. However, we notice that the linear sizes of the EGO cloud cores are very similar among the nine group-I objects that have the linear sizes most reliably determined. Thus, we adopt the average linear size of $\overline{S}=0.036$\,pc of the nine group-I objects for the group-III optically thin objects, so that we can estimate the mass and density for them. It also allows us to estimate a typical true methanol line opacity from the beam averaged line opacity via $\tau=\tau/f$ for each object, as mentioned earlier in Sect.~\ref{para_dengeneracy}.

The resulting physical parameters for all 29 EGO cloud cores (groups-I, II and III in Table~\ref{tab_ch3oh}) are given in Table~\ref{tab_ldsmn}. Using the average small-scale methanol abundance, the cloud core masses are found to range from a few tenth to a few tens solar masses and the densities vary from a few $10^5$ to higher than about $10^7$\,cm$^{-3}$. Note that the methanol abundances of individual EGO cloud cores can be different from the average value by an order of magnitude. Thus, the core masses and gas densities estimated with the average abundance can be uncertain by one order of magnitude, too. Combining with their typical linear size of about 0.036\,pc, the physical properties of the EGO cloud cores are typical for dense molecular cloud cores in our Galaxy \citep[see e.g., ][]{herb2009}.

\section{Discussion}
\label{discuss}

\subsection{Chemical and physical properties of EGO cloud cores}
\label{EGOscloudcore}

\subsubsection{Organic chemistry in EGO cloud cores}
\label{discuss_methanol}

According to the recent astrochemical modeling of star forming cloud collapse and warm-up processes by \citet{garr2006,garr2008} and \citet{garr2013}, the gas phase methanol is produced in two distinct steps: In the first step, it is produced by gas phase reaction after the evaporation of H$_2$CO at a gas temperature of $\sim 40$\,K. Then, the gas phase methanol abundance drops due to accretion back onto the icy grain surfaces. In the second step, it is directly evaporated from the grain surfaces at $\sim 90$\,K. The first step starts at $\sim 2\times 10^4$\,yr for the massive fast warm-up models or $\sim 8\times 10^4$\,yr for the intermediate mass medium warm-up models \citep[see e.g., the Figs. 1 and 2 of][]{garr2013} and the methanol abundance can mount up to a few $10^{-9}$. The second step occurs at  $\sim 3.3\times 10^4$\,yr for the massive fast warm-up models or $\sim 1.3\times 10^5$\,yr for the intermediate mass medium warm-up models and the methanol abundance can reach a few $10^{-5}$. 

To compare the observed methanol abundances in the small EGO cloud cores with the model values, we need to know how to convert the H$_2$ column density in a large beam into that in compact cloud cores. Because of the lack of observational information, we resort to empirical radial density profile models of clouds. \citet{berg2007} reviewed the density structure of three individual prestellar cores that were determined from the observations of dust extinction at near infrared (Barnard\,68), dust absorption at mid-infrared ($\rho$\,Oph) and dust emission at (sub)millimeter (L1544), respectively. The three cores show similar two-component density structures: a dense core with a density gradient flatter than $r^{-1}$ (we assume $r^0$ here) within a radius $r_1<\sim 5000$\,AU and an envelope around the core with a power law density profile close to $r^{-2}$ within a radius of about 10 times of $r_1$. At larger radii beyond $10r_1$, we can adopt the power law density profile of $r^{-1}$ which is consistent with the Larson's scaling law \citep{lars1981,solo1987}. We consider a typical EGO cloud methanol core diameter of 0.036\,pc at a typical distance of 4\,kpc, and assume it has a similar density structure as above and fills the whole telescope beam of $65.5\arcsec$ in the C$^{18}$O\,1-0 observations of \citet{chen2010} (corresponding to a diameter of about 1.27\,pc for an EGO at this distance). Then, integrating the density profile within the methanol core and within the whole telescope beam shows that the average H$_2$ column density in the methanol core should be higher than in the large beam by a factor of $\sim 4$. Because the typical filling factor of the methanol cores in our EGO clouds is $6.3\times 10^{-3}$ (average among the group-I objects), the methanol abundances in the cores should be higher than the beam averaged ones by a factor of about 40 on average, reaching $1.28\times 10^{-7}$ (for all 29 EGOs). Similarly, the average abundances of the other three species in the cores become 
$5.44\times 10^{-8}$ for CH$_3$OCH$_3$, 
$3.02\times 10^{-7}$ for HCOOCH$_3$ and 
$1.58\times 10^{-8}$ for CH$_3$CH$_2$CN. 

Comparing these abundances and their ratios with the two gaseous methanol producing stages (early stage around 40\,K and hot core stage warmer than 90\,K) in the three-phase grain chemistry cloud warm-up models of \citet[][Fig. 1-3]{garr2013}, we argue that the early stage can be ruled out for the EGO cloud cores. The models predict either several orders of magnitude lower abundances or orders of magnitude different abundance ratios, as compared to the observed values.
 
The later hot core stage also does not completely fit to our data. For example, in the most probable massive fast warm-up model case, the model predicts about two orders of magnitude higher methanol abundances (about $10^{-5}$) in this stage than we obtained for EGO cloud cores (about $10^{-7}$), although the abundances of CH$_3$OCH$_3$ and HCOOCH$_3$ are correctly predicted by the model. Moreover, the model methanol abundances are two orders of magnitude higher than that of the other two O-bearing species, while the observed abundance ratios are not far from unity.

The only phase of the models during which both the observed abundances and abundance ratios in EGO cloud cores can be correctly reproduced is the short beginning time of the hot core stage in the massive fast warm-up model (the onset of the hot core stage). It is the time when the methanol molecules are evaporating from the icy grain surfaces at a gas temperature of about 90\,K and a chemical age of about $3.5\times10^4$\,yr. This conclusion does not contradict with previous claims that EGOs are massive star forming clouds in early stages \citep[before the hot core stage, see e.g.,][]{cyga2008,paron2012,kend2013}. However, this phase is so short, covering only a narrow gas temperature range of about $95-110$\,K and a similarly narrow range of about $3.1-3.7\times 10^4$\,yr of chemical age, that it is difficult to explain why the majority of the EGO cloud cores fall into so narrow time gap.
 
In the above discussions, the typical kinetic temperature in the EGO cloud cores is required to be about 100\,K, which is significantly higher than the average methanol rotation temperature of 44\,K (ranging from 19-81\,K). To investigate the reason of this difference, we first compute the critical densities of the observed CH$_3$OH lines at temperatures of 44 and 100\,K, which yields $n_{cr}\sim 10^6$ cm$^{-3}$ at both temperatures. The gas densities estimated for the small cloud cores in Sect.~\ref{massdensity} range from $10^5$ to $>10^7$\,cm$^{-3}$, which indicates that the involved methanol lines in the cloud cores may be only marginally thermalised. Furthermore, methanol is a well known maser molecule (easy to have non-thermal excitation). According to \citet{crag1992}, the level populations are determined by collisional excitation and selective cascade in regions without strong dust emission (e.g., class\,I maser regions in outflow shocks), while in regions with prominent dust emission (e.g., class\,II maser regions in hot cores), infrared radiation pumping will dominate. In both cases, the different K-ladders will be non-thermally populated in the typical interstellar conditions, which results in masers in some transitions. Therefore, non-thermal excitation of thermal methanol lines are also expected. It is known that EGOs often possess both classes of methanol masers \citep[see][]{chen2009} and thermal methanol line emission can simultaneously appear in both outflow and core regions of an EGO \citep[see e.g.,][]{cyga2011a}. The fact from Paper\,I that the methanol line widths do not correlate with that of either SiO\,6-5 lines or H$^{13}$CO$^+$\,3-2 lines also supports that both hot core and outflow shock regions may contribute to the observed thermal methanol lines. The contribution from shock regions, as shown by \citet{bach1995} and \citet{saka2008,sakai2010a} in some dark clouds, tends to produce a rotation temperature lower than gas kinetic temperature. For the contribution from the hot cores, as investigated by \citet{crag1992}, the $K=2$ ladder of A-type methanol molecules tend to be overpopulated relative to the $K=3$ ladder, and thus the methanol $J_K=J_3-J_2$ rotation temperature found in this work is also expected to be lower than the kinetic temperature. Thus, the sub-thermality, effects of radiation pumping and composite contributions from both shock and hot core regions, all support that higher gas kinetic temperatures than the methanol rotation temperatures are possible.

According to \citet{garr2006,garr2008} and \citet{garr2013}, dimethyl ether (CH$_3$OCH$_3$) and methyl formate (HCOOCH$_3$) should be also produced through similar gas phase reaction networks initiated by the evaporation of either H$_2$CO in the early warm-up phase or  CH$_3$OH in the later hot core phase. For ethyl cyanide (CH$_3$CH$_2$CN), it could be also produced by gas phase reactions triggered by the evaporation of either HCN in the early warm-up phase or CH$_3$CN in the later hot core phase. Hence, the abundances of CH$_3$OH, CH$_3$OCH$_3$ and HCOOCH$_3$ and possibly CH$_3$CH$_2$CN are expected to be correlated.

\begin{figure}
 \centering
 \includegraphics[angle=0,height=6cm,width=8cm]{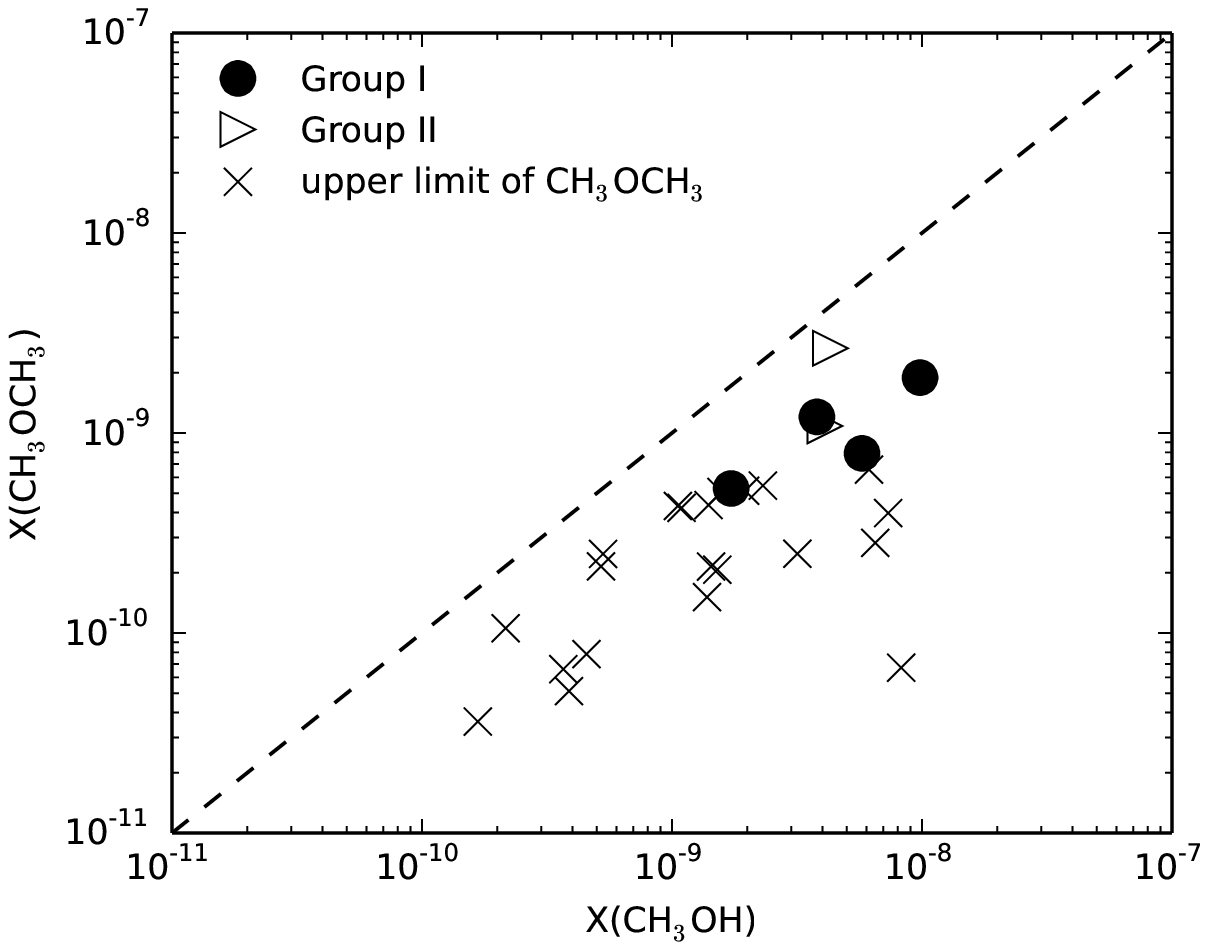}
 \includegraphics[angle=0,height=6cm,width=8cm]{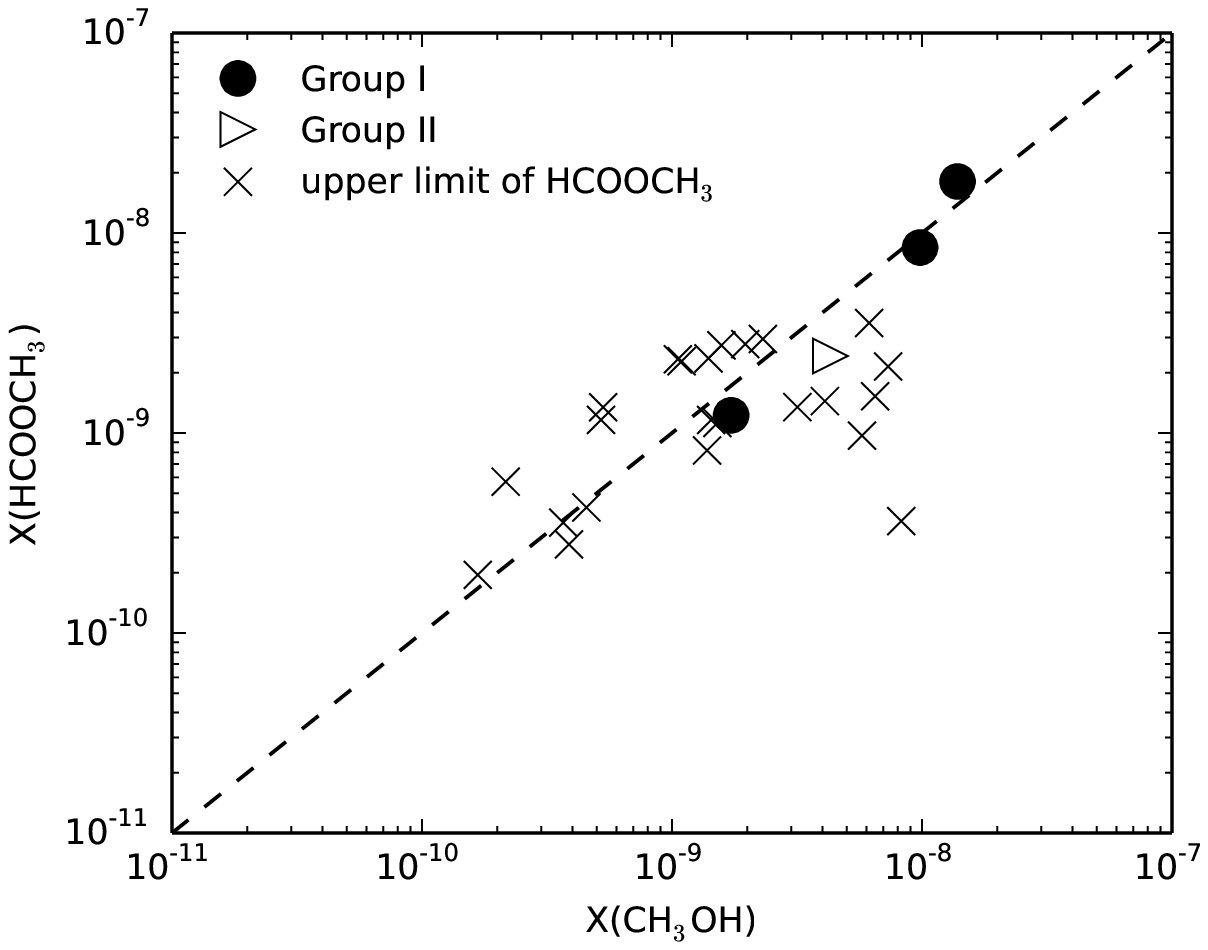}
 \includegraphics[angle=0,height=6cm,width=8cm]{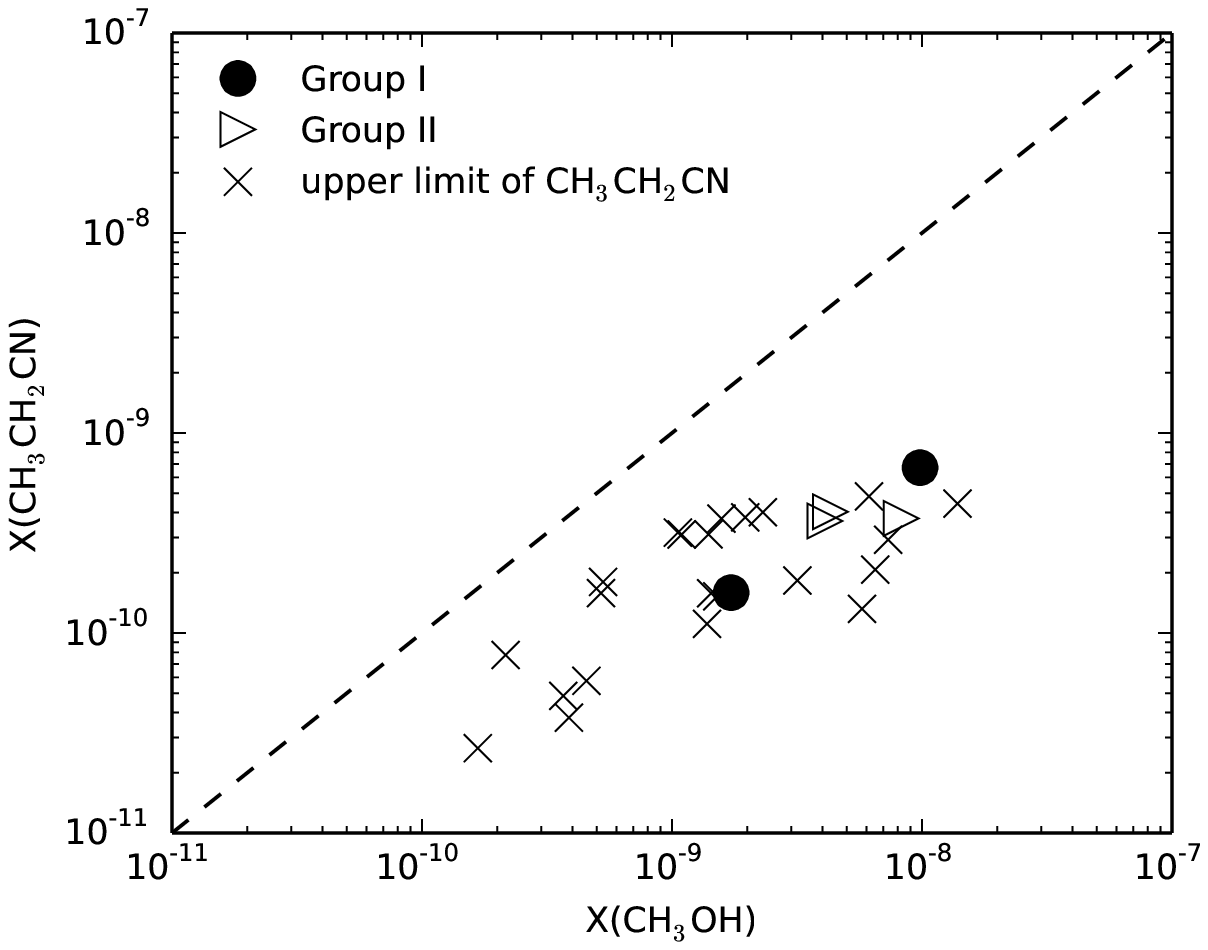}
 \caption{The relations of abundances of the three complex molecules with that of methanol. The group-II objects only have lower limits of the CH$_3$OH abundance (right triangles) because of the optically thick lines. The upper limits of non-detections (crosses) of CH$_3$OCH$_3$, HCOOCH$_3$ and CH$_3$CH$_2$CN are estimated using baseline noise and assumed line width and rotational temperature derived from methanol. The dashed lines show the $Y=X$ relationship to help visual comparison. Error bars are not plotted, because they are usually large for methanol and seriously underestimated for the other three species.}
 \label{fig_ab}
\end{figure}

To verify this, we compare the observed beam averaged abundances of CH$_3$OCH$_3$, HCOOCH$_3$ and CH$_3$CH$_2$CN with that of methanol in Fig.~\ref{fig_ab}. We draw a dashed diagonal line in each panel to show the $Y=X$ linear relationship. It is clear that the data distribution trends of groups-I and II objects (black dots and right triangles) are roughly parallel to the dashed lines in all three panels, indicating linear correlations among the abundances of all four organic species. This result agrees with the above theoretical expectation.

Note that we do not show error bars in Fig.~\ref{fig_ab} and the trends in these figures are very tentative, because the error bars of methanol abundances of group-I objects are usually very large (uncertain by a factor of about 2.2 on average) and that of the group-II objects are unavailable (lower limits). The error bars of the other three species should be even larger but are underestimated (due to fixing the rotational temperature and filling factor in the PD analysis, which has resulted in a too small typical error of about 20 per cent).

Because of the linear correlations in Fig.~\ref{fig_ab}, it is reasonable to compute an average abundance ratio between each pair of the four organic species using the beam averaged abundances in Tables~\ref{tab_ch3oh} and \ref{tab_other3}. Then, we combine the ratios and arbitrarily normalize the methanol abundance to an integer number of 200 to derive the abundance ratios [CH$_3$OH : CH$_3$OCH$_3$ : HCOOCH$_3$ : CH$_3$CH$_2$CN] $\approx 200 : 61 : 172 : 15$.  

Therefore, were the chemical modeling works of \citet{garr2006,garr2008} and \citet{garr2013} reliable, only the brief onset phase of the hot core stage can reproduce both the observed small-scale organic abundances and their ratios. At this stage, the gas temperature in the core is about 100\,K, while the methanol $J_K=J_3-J_2$ rotation temperature is only about 44\,K. All four organic species have comparable abundances on the orders of $10^{-8}$ to $10^{-7}$, and their abundances are possibly linearly correlated. However, the time length of the onset phase of the hot core stage is too short to get all the EGO cloud cores explained.

\subsubsection{Parameter correlations}
\label{Luminosity}

\begin{figure*}
 \centering
 \includegraphics[angle=0,height=4cm,width=5.5cm]{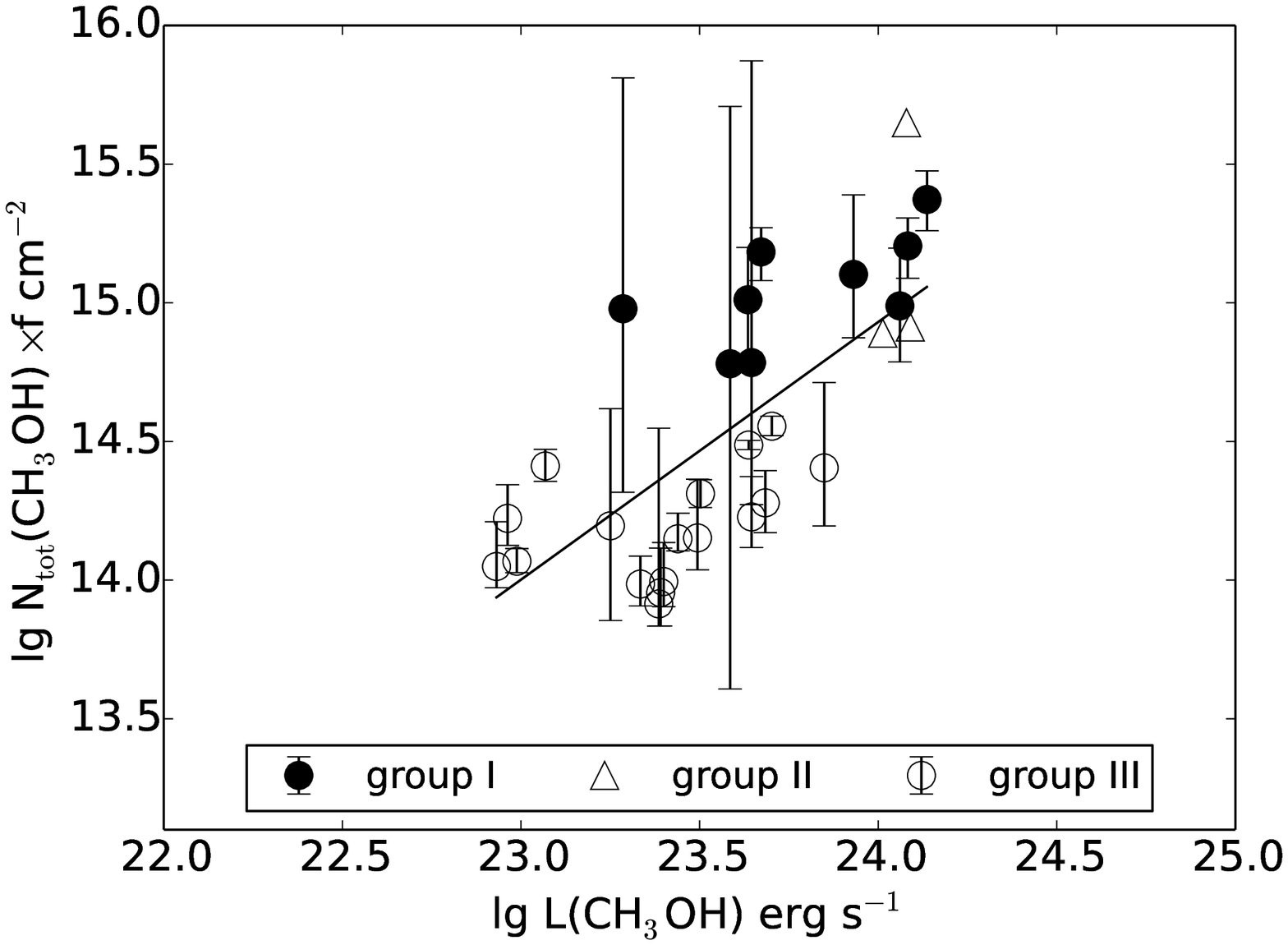}
 \includegraphics[angle=0,height=4cm,width=5.5cm]{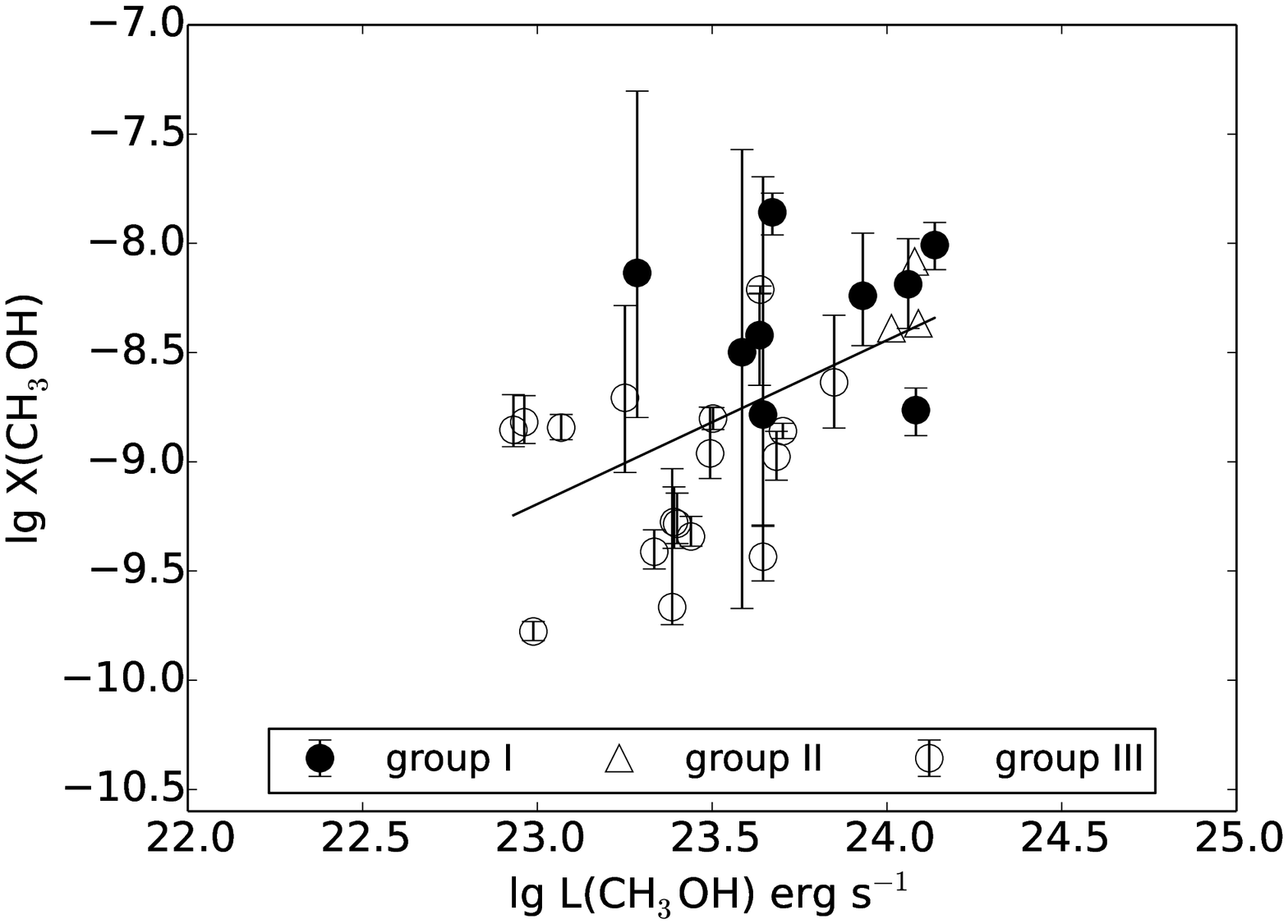}\\
 \includegraphics[angle=0,height=4cm,width=5.5cm]{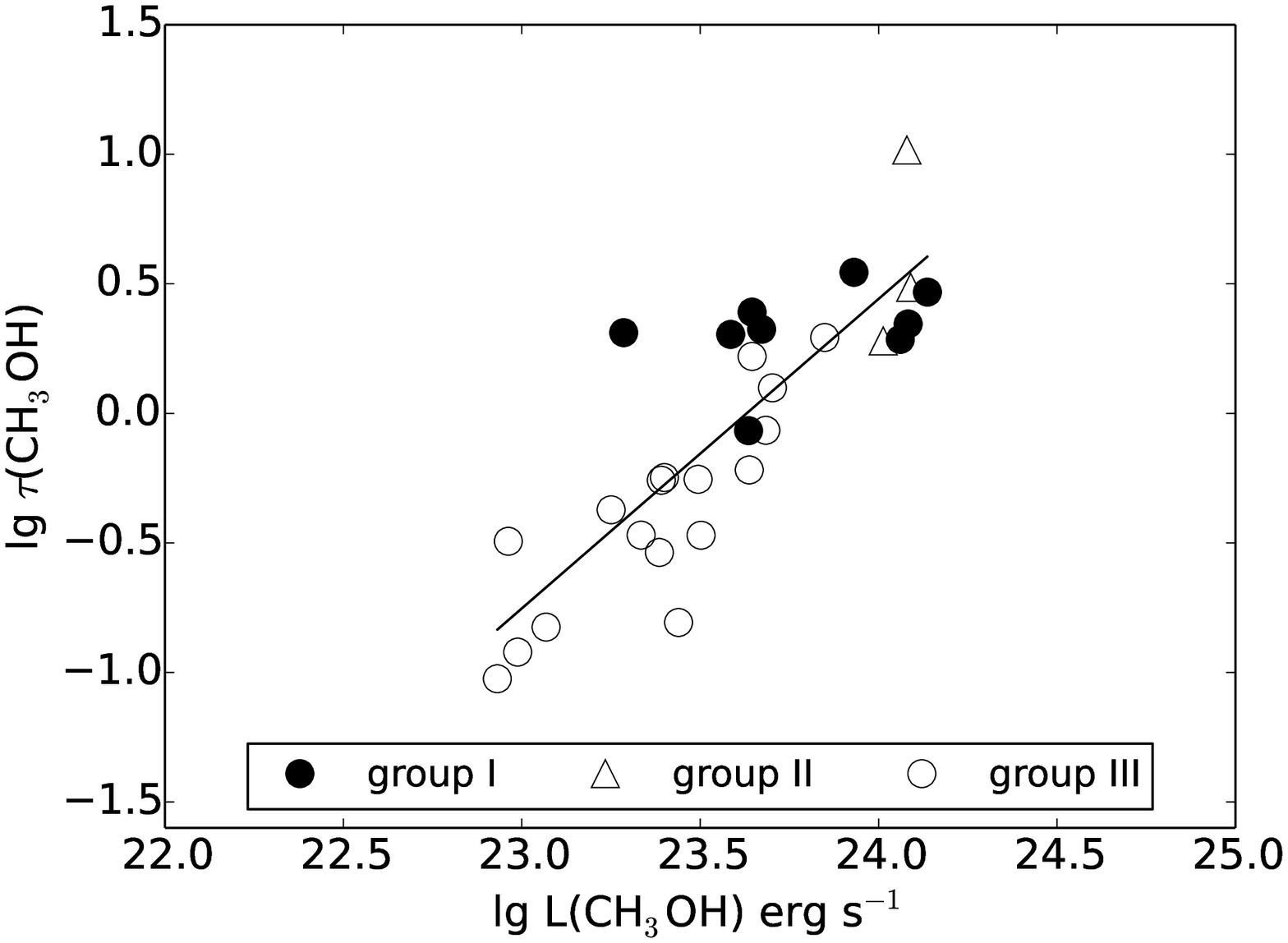}
 \includegraphics[angle=0,height=4cm,width=5.5cm]{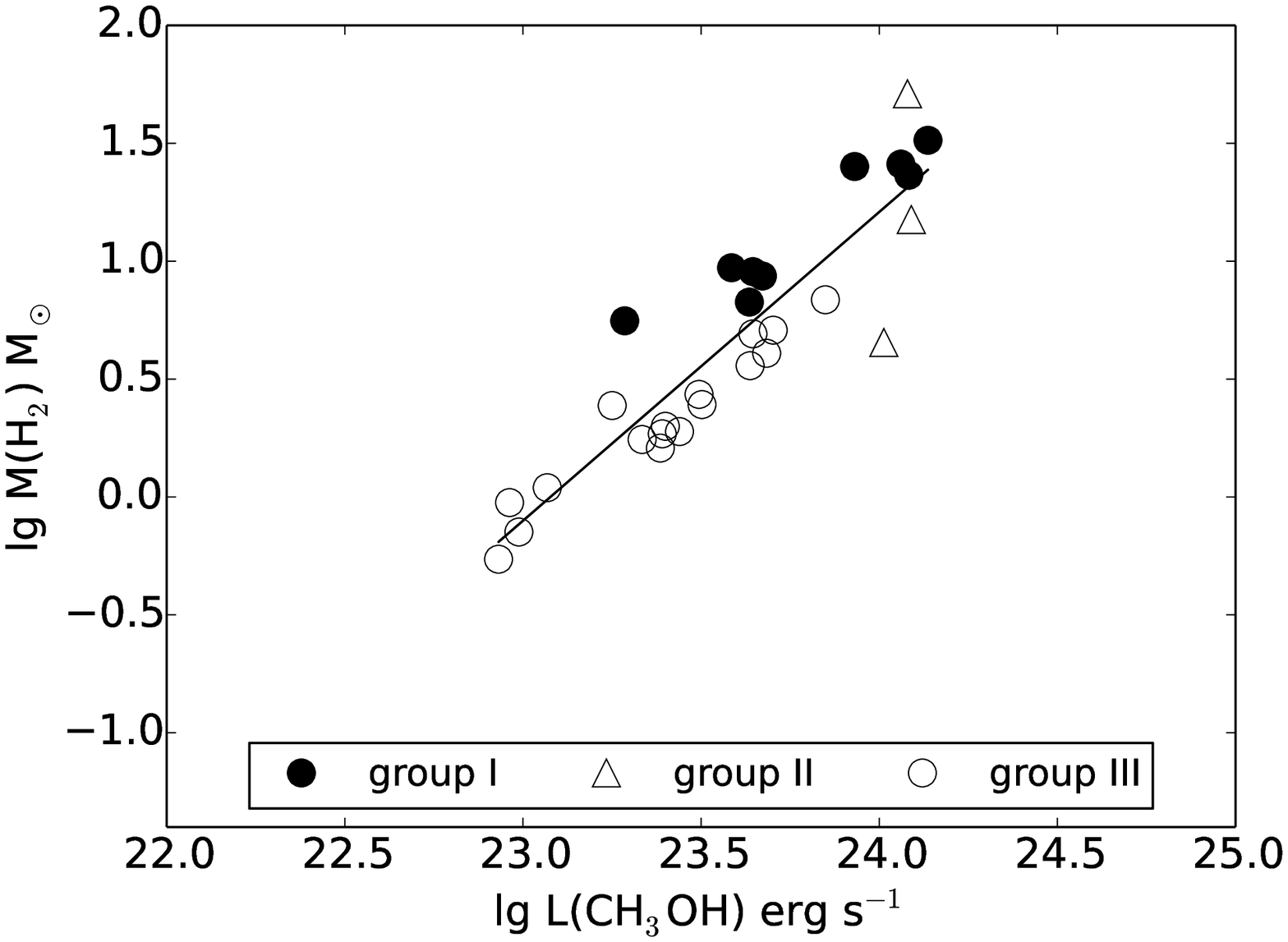}
 \includegraphics[angle=0,height=4cm,width=5.5cm]{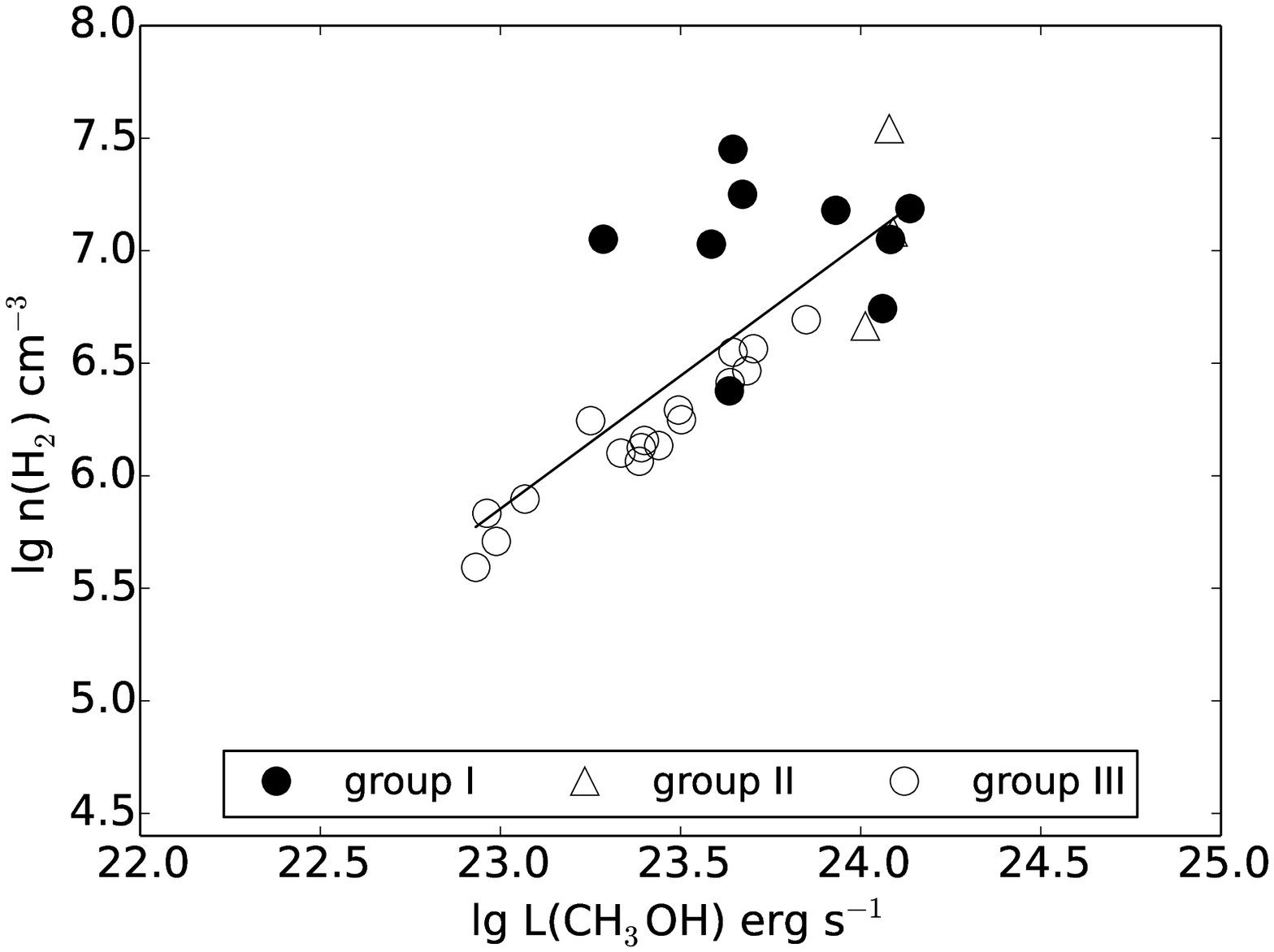}\\
 \includegraphics[angle=0,height=4cm,width=5.5cm]{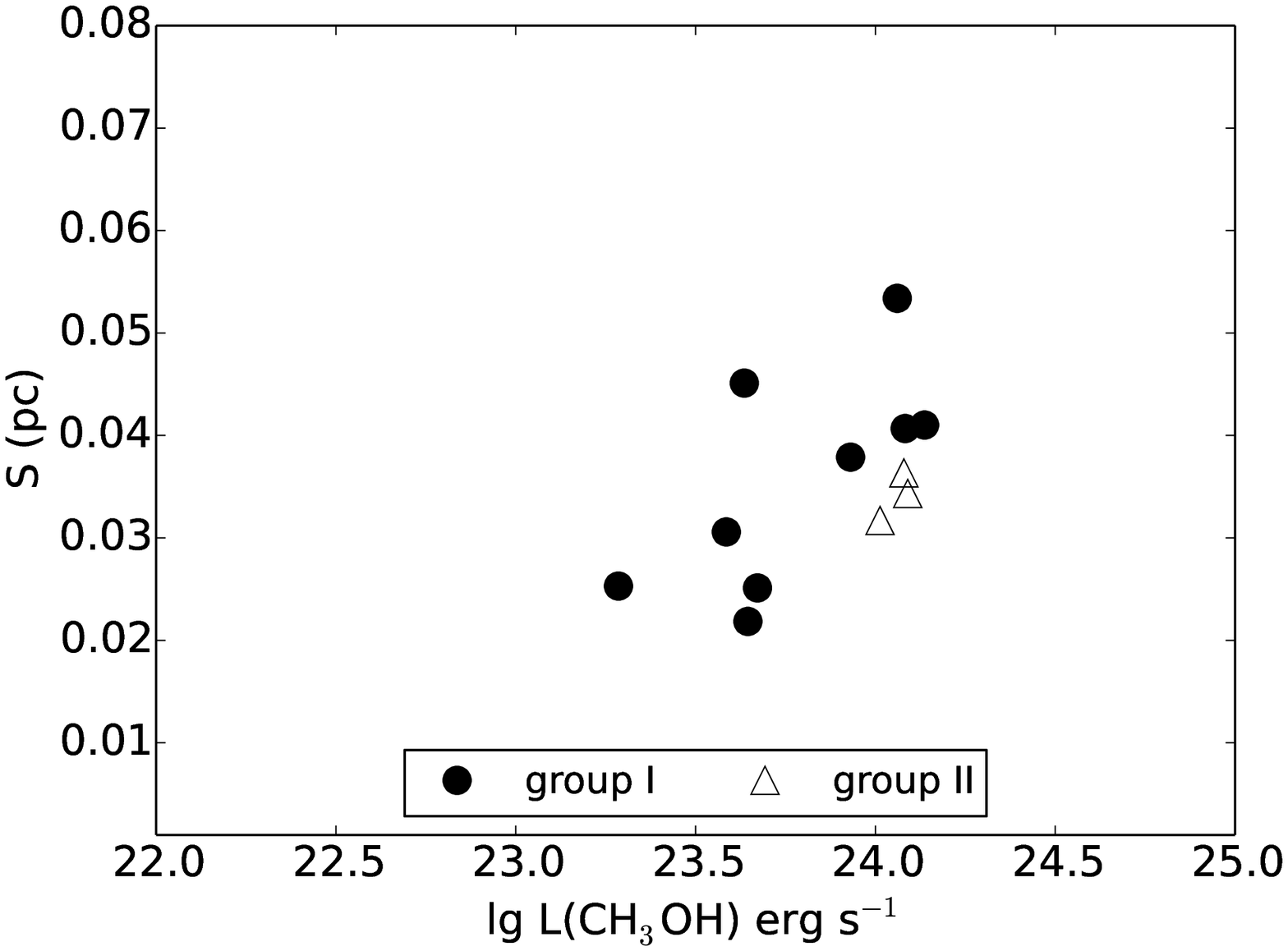}
 \includegraphics[angle=0,height=4cm,width=5.5cm]{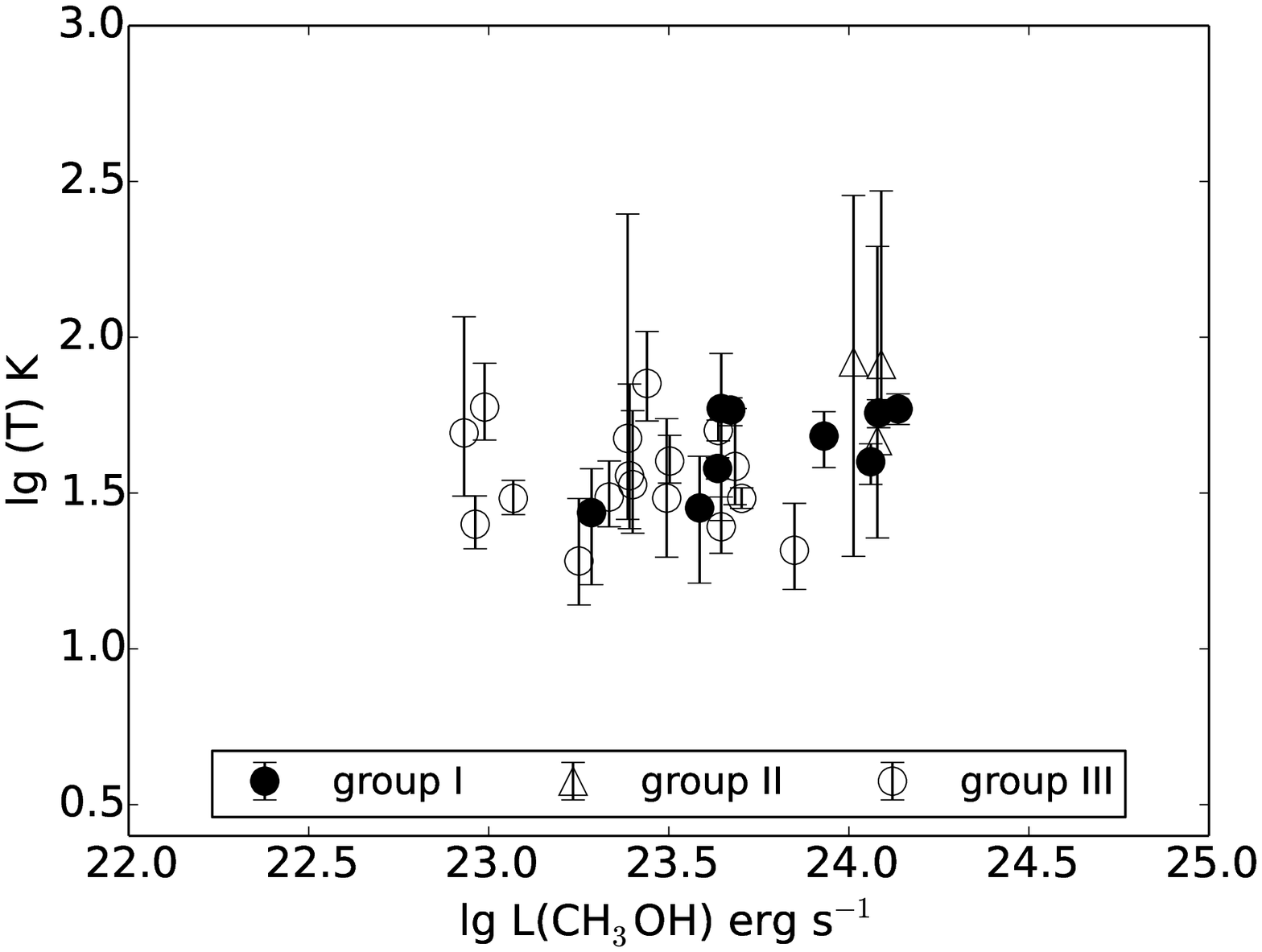}
 \caption{The correlations with methanol line luminosities of other parameters: column density $N_{\rm tot}$, relative abundances $X$ and optical depths $\tau$ of methanol, total gas masses $M$, densities $n$, linear sizes $S$, and methanol rotation temperatures $T_{\rm rot}$ of the cloud cores. Error bars are shown only for $N_{\rm tot}$, $X$ and $T$, because that of the other quantities are very uncertain. The straight lines in the first five panels are the log-linear fits represented by Eqs.~(\ref{ncl}-\ref{nl}).}
 \label{fig_corr}
\end{figure*}
\begin{figure}
 \includegraphics[angle=0,height=6cm,width=8cm]{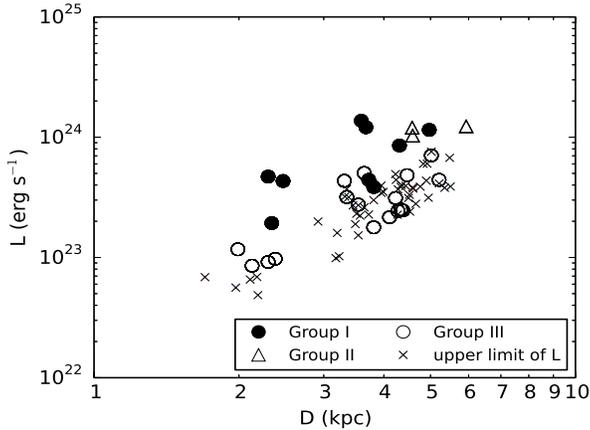}
 \caption{Distance dependence of methanol line luminosities of our EGO samples.}
 \label{fig_LD}
\end{figure}
To show the correlations between the physical parameters of the EGO cloud cores, we compare them with the CH$_3$OH line luminosities of the same objects from Paper\,I in Fig.~\ref{fig_corr}. The beam averaged CH$_3$OH column densities $N_{tot}\times f$ and abundances $X$, methanol line opacities $\tau$, gas mass $M$ and gas density $n$ (assuming a constant methanol abundance) all show correlations with the methanol line luminosities in the logarithmic scales. We fit log-linear relations to these plots, which yields 
\begin{eqnarray}
\lg (N_{tot}\times f) =0.93\lg L-7.37  &(R=0.68, \sigma=0.20),
\label{ncl}
\end{eqnarray}
\begin{eqnarray}
\lg X=0.75\lg L-26.44 &(R=0.49, \sigma=0.27),  
\label{xl}
\end{eqnarray}
\begin{eqnarray}
\lg \tau=1.19\lg L-28.24 &(R=0.84, \sigma=0.15),
\label{tl}
\end{eqnarray}
\begin{eqnarray}
\lg M=1.31\lg L-30.22 &(R=0.91, \sigma=0.12),
\label{ml}
\end{eqnarray}
\begin{eqnarray}
\lg n=1.18\lg L-21.32 &(R=0.79, \sigma=0.18),
\label{nl}
\end{eqnarray}
where $R$ is the linear correlation coefficient and $\sigma$ is the standard error of the estimate for the logarithmic quantities. 

In the bottom left panel of the figure, reliable linear source sizes are available for the groups-I and II objects whose methanol line opacities are close to unity or higher. The points are scattered in a small range of sizes ($0.02-0.06$\,pc) in the plot and show a dependence on line luminosities. The average cloud core size among group-I objects is about 0.036\,pc.

In the bottom right panel, the methanol rotation temperatures are crowed around an average value of 44\,K and show no obvious trend with the methanol luminosities. The three methanol opacity groups (the different symbols) also do not show systematic difference.

One may worry about whether some of these correlations purely result from the uncertainties in the kinematic distances. We point out that only four of the quantities involve distance: $L\propto D^2$,  $M\propto D^2$, $n\propto D^{-1}$, and $S\propto D$, which correspond to correlations $\lg M\propto \lg L$, $\lg n\propto -0.5\lg L$, and $S\propto \sqrt{10}^{\lg L}$. However, the correlations shown in Fig.~\ref{fig_corr} or in Eqs.~(\ref{ml}) and (\ref{nl}) are quite different from the above ones, which demonstrates that the correlations are not due to distance effects. Furthermore, we check the relation between the methanol line luminosities and the distances in Fig.~\ref{fig_LD}. Also shown in the figure are the methanol line luminosity upper limits of those objects without methanol lines but with H$^{13}$CO$^+$\,3-2 line detected in Paper\,I. Although the methanol line luminosities are clearly correlated with distances in the figure, we argue that this is very possibly merely a selection effect. This is because the more luminous objects are more easily to get detected at greater distances. The line luminosity upper limits of methanol non-detections (crosses in the figure) are distributed immediately below the detected objects, indicating that there might exist many weak but distant methanol emitters in our samples, but they are not detected simply due to our limited survey depth.

All in all, we can get an overall picture upon the physical nature of the EGO cloud cores. They are cloud cores of very different masses (assuming a constant methanol abundance), with more massive ones more distant (selection effect). It is not clear if they are cloud cores born in similar environments but in different evolutionary stages or cloud cores produced in very different kinds of interstellar cloud. Nevertheless, they share very uniform properties: the methanol line opacities, core masses, gas densities, and possibly methanol column densities and abundances log-linearly correlate with the methanol line luminosities, while these cores have similar linear sizes of about 0.036\,pc and similar methanol $J_3-J_2$ rotation temperatures of about 44\,K. To express some of the correlations in a more general way that is independent of the specific molecular transitions used, we can merge Eqs.~(\ref{ml}) and (\ref{nl}) into one
\begin{equation}
 n=5.0\times 10^{7}(M/10^2M_\odot)^{0.9}\,{\rm cm}^{-3}.
\label{nm}
\end{equation}
It demonstrates that the gas densities in the cores are nearly linearly correlated with the core masses, which is in accord with the fact that most of the cores have similar core sizes.

\subsection{Comparing EGO cloud cores with other observed cloud cores}
\label{compare_other_cloud}

\subsubsection{Beam-averaged column densities}
\label{comparecolumn}

To compare with other works, we stress that, in determining the column density of complex molecules, not only the different telescope beam sizes and observational errors such as telescope pointing error, flux calibration error, may cause trouble to the comparison, but the different choices of partition functions are also an issue. For example, \citet{blak1987} pointed out that the low-lying vibrational states of methanol may be significantly populated even when the gas temperature is as low as 100-200\,K. In this work, we always use the partition functions from the CDMS/JPL databases that have the lowest vibrational levels included. Therefore, one should keep in mind that the comparison of both column densities and abundances in this section has not fully taken into account all sources of uncertainty.

The beam averaged CH$_3$OH column densities of our EGO cloud cores are in the range of $\sim 10^{13}$ to $\sim 10^{15}$ cm$^{-2}$, which are very similar to that found by \citet{chen2013b} toward a newer sample of EGOs identified from {\it GLIMPSE\,II} survey. They are also on the same orders of magnitude as that obtained toward other cloud cores in literature. For example, \citet{sakai2010a} found CH$_3$OH column densities from $\sim$ 2.9$\times 10^{13}$ to $\sim 9.7\times 10^{14}$\,cm$^{-2}$ toward some massive clumps in early evolutionary stages of high-mass star formation regions; \citet{beut2007} obtained an average CH$_3$OH column density of $8.2\times 10^{13}$\,cm$^{-2}$ for massive infrared dark clouds; \citet{sanh2013} derived a CH$_3$OH column density of  about $1.1\times 10^{14}$ in the very massive quiescent cold clumps (MM1) in IRDC G028.23-00.19.

\subsubsection{ Abundances and abundance ratios}
\label{compareabun}
\begin{table*}
 \centering
 \begin{minipage}{160mm}  
  \caption{{\bf Comparison of} relative abundance ratios of EGO cloud cores with other types of clouds.}
  \label{tab_clouds}
  \begin{tabular}{@{}l@{ }l@{ }l@{ }l@{ }lll@{}ll@{}}
  \hline
Objects                   & Abundance ratios     & N$_{\rm object}$ & X(CH$_3$OH) & Telescope & $\lambda$(mm) & Beam($''$)& Q$^a$ & Ref$^{*}$  \\
                          & (1):(2):(3):(4)      & (2,3,4)   & $\times 10^{-9}$.  &           &               &    &   &     \\
\hline
EGOs                      & 200: 61 : 172 :  15  & 6,4,5     &   3.20 & AROSMT10m  &1.2        & 29         &  V & this work \\

\multicolumn{9}{|l|}{--- GC clouds---}\\
GC clouds                 & 200:  8 :   8 : ---  & 40,40,-   &  350.0 & IRAM30m    &1.3,2,3    & 12,17,24   &  R & Requ06 \\
SgrB2                     & 200:   5 :   3 : --- & 1,1-      &   20.0 & BTL7m      &2.1,4.2    & 90,174     &  N & Cumm86 \\
SgrB2(N)                  & 200:   7 :   5 :  10 & 1,1,1     &   20.0 & NRAO12m    &1.3        & 25         &  N & Iked01 \\
SgrB2(N) core$^b$         & 200:   3 :  1 :  0.6 & 1,1,1     &  200.0 & SEST15m    &1.3        & 23         &  V & Numm00 \\
SgrB2(M) core$^b$         & 200:   2 :  1 :  0.1 & 1,1,1     &  200.0 & SEST15m    &1.3        & 23         &  V & Numm00 \\

\multicolumn{9}{|l|}{--- Massive cloud cores---}\\
Massive hot cores         & 200:  81 :  18 :   2 & 7,7,7     & 1900.0 & JCMT15m    &0.8,1.3    & 14,21      &  R & Biss07 \\
                          &                      &           &        & IRAM30m    &1,2,3      & 12,17,24   &             \\
G327.3-0.6                & 200:  86 : 143 :  23 &           &   70.0 & SEST15m    &1.3,2.6    & 22,48      &  R & Gibb00 \\
hot core G34.3+0.2        & 200:  22 :  11 :   2 & 1,1,1     &   90.0 & NRAO12m    &1.3        & 25         &  N & Iked01 \\
                          &                      &           &        & NRO45m     &3          & 16         &             \\
                          &                      &           &        & SEST5m     &1.3,3      & 22,51      &             \\
hot core G327.3-0.6       & 200:  60 : --- : --- & 1,-,-     &  100.0 & SEST5m     &1.3,3      & 22,51      &  N & Iked01 \\
hot core W51 e1/e2        & 200: --- :  20 : --- & -,1,-     &  300.0 & NRAO12m    &1.3        & 25         &  N & Iked01 \\
                          &                      &           &        & NRO45m     &3          & 16         &             \\
G34.26+0.15(SE)hotcore    & 200:  27 :  21 :   2 & 1,1,1     &  640.0 & BIMA6.1m$\times$4 &2.7,3.4    & 106,132    &  N & Mook07 \\
G34.26+0.15(NE)hotcore    & 200:  33 :  17 :   1 & 1,1,1     &  850.0 & BIMA6.1m$\times$4 &2.7,3.4    & 106,132    &  N & Mook07 \\

\multicolumn{9}{|l|}{--- OMC ---}\\
OMC NW Plateau            & 200: 120 : 120 :  30 & 1,1,1     &   10.0 & JCMT15m    &0.8        & 13.7       &  R & Sutt95 \\
Orion SE Plateau          & 200:   3 :  10 :   4 & 1,1,1     &  220.0 & JCMT15m    &0.8        & 13.7       &  R & Sutt95 \\
Compact Ridge in OMC-1    & 200:  30 :  15 : --- & 1,1,-     &  120.0 & OVRO40m    &1.3        & 30.0       &  V & Blak87 \\
OMC Compact Ridge         & 200:  10 :  15 :   3 & 1,1,1     &  400.0 & JCMT15m    &0.8        & 13.7       &  R & Sutt95 \\
OMC Hot Core              & 200:  11 :  20 :   4 & 1,1,1     &  140.0 & JCMT15m    &0.8        & 13.7       &  R & Sutt95 \\
Orion hot core            & 200: --- :  18 :   4 & -,1,-     & 1000.0 & NRAO12m    &1.3        & 25         &  N & Iked01 \\
                          &                      &           &        & NRO45m     &3          & 16         &             \\

\multicolumn{9}{|l|}{--- hot corinos ---}\\
IRAS 16293-2422           & 200:  480: 340 :  24 & 1,1,1     &  100.0 & IRAM30m    &1,2,3      & 12,17,24   &  N & Bott07 \\
NGC1333-IRAS2A            & 200:  20 : 447 :  67 & 1,1,1     &  300.0 & IRAM30m    &1,2,3      & 12,17,24   &  N & Bott07 \\
NGC1333-IRAS4B            & 200: 343 : 314 : 214 & 1,1,1     &  700.0 & IRAM30m    &1,2,3      & 12,17,24   &  N & Bott07 \\
\hline
\end{tabular}\\ [1mm]
Note: The columns are
Object names or types;
Abundance ratios (1):(2):(3):(4) = X(CH$_3$OH) : X(CH$_3$OCH$_3$) : X(HCOOCH$_3$) : X(CH$_3$CH$_2$CN), with X(CH$_3$OH) normalized to 200;
Number of objects averaged together for the abundance ratios (only for EGOs from this work, massive hot cores from Biss07, and GC clouds from Requ06);
{\bf Beam averaged} CH$_3$OH abundance;
Telescope(s) and diameter(s);
Observation wavelength;
Telescope beam; Notes on used partition function (Q, see footnote {\em a} below);
Reference codes.\\
{\em a}:
      `V': Partition function with low-lying vibrational states included;
      `R': ground vibrational state only;
      `N': not mentioned in the literature.\\
{\em b}: The methanol lines were fitted with a two-component model (core and halo) in this literature. We use the core abundances because the weaker emission of the other three species are very possibly dominated by the hot and dense core component.\\
{\rm *}: References:
          Numm00 = \citet{numm2000};
          Biss07 = \citet{biss2007};
          Blak87 = \citet{blak1987};
          Bott07 = \citet{bott2007};
          Cumm86 = \citet{cumm1986};
          Gibb00 = \citet{gibb2000};
          Iked01 = \citet{iked2001};
          Mook07 = \citet{mook2007};
          Requ06 = \citet{requ2006};
          Sutt95 = \citet{sutt1995}.
\end{minipage}

\end{table*}

To compare the organic chemical properties of the EGO cloud cores with other types of known cloud cores, we collect in Table~\ref{tab_clouds} the abundances and abundance ratios of the same four species in some representative clouds of other types such as massive hot cores, hot corinos and clouds formed at the dynamical environments near the Galactic Center. Also collected are information about the telescopes, observation wavelengths, and methanol partition function used in the literature works, which shall help us check whether they affect our comparison.

Entries in Table~\ref{tab_clouds} are grouped according to their cloud types: EGOs, GC clouds (including Sgr\,B2), massive cloud cores (including the subgroup of OMC) and hot corinos. Particularly, for the EGOs from this work, some massive hot cores from \citet{biss2007}, and some GC clouds from \citet{requ2006}, we only use the average abundances and abundance ratios from the original literature. They are listed near the top of the table and we give the number of objects averaged together. For the other individual objects, they are sorted in increasing methanol abundances in each (sub)group. 

\begin{figure}
 \centering
 \includegraphics[angle=0,height=11.5cm,width=8cm]{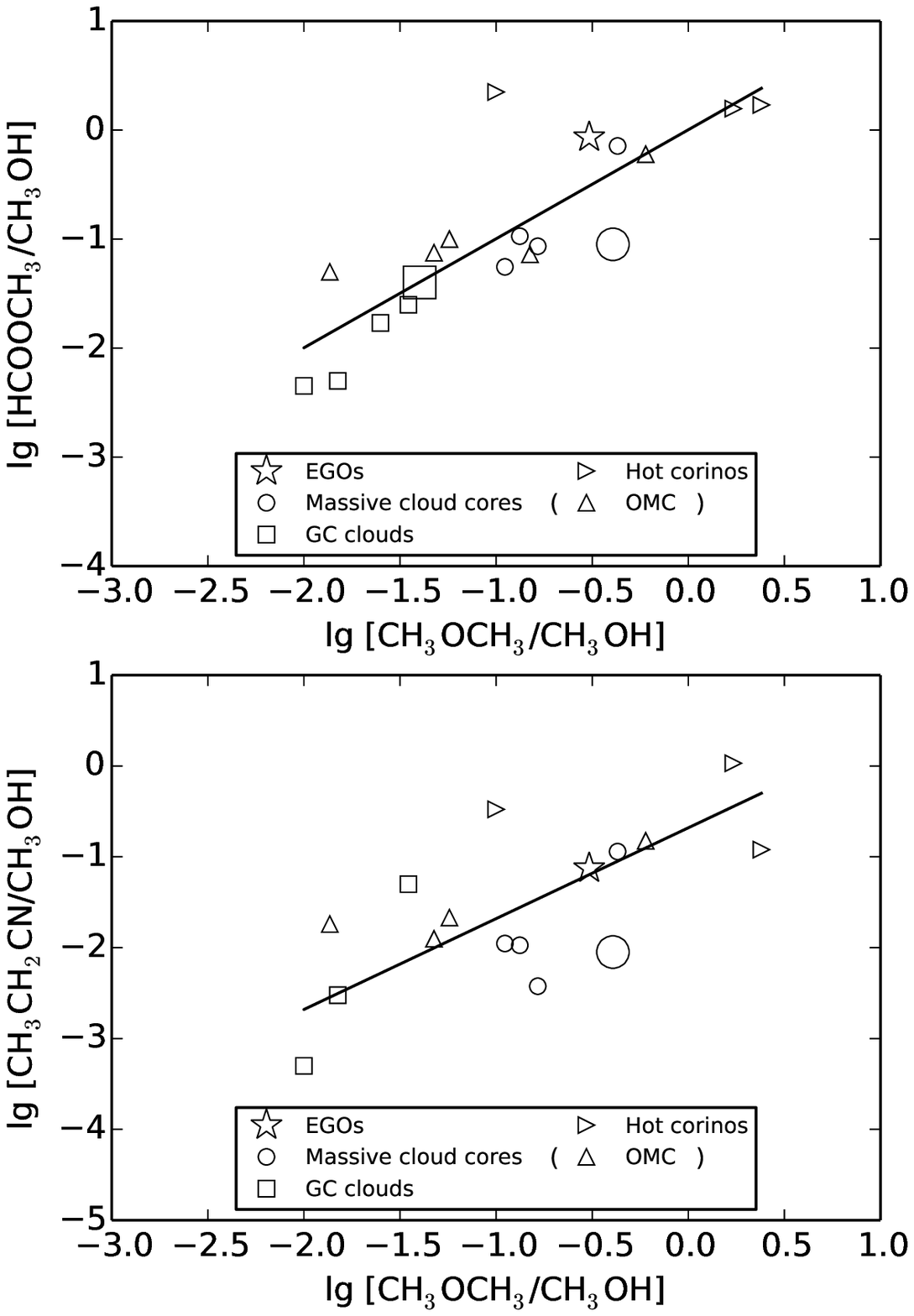}
 \caption{The correlation of organic abundance ratios among the different types of cloud cores. The three bigger symbols ( a star, a circle and a square) are for the average abundance ratios of EGOs, some massive hot cores and some GC clouds, respectively (see the details in Table~\ref{tab_clouds}). The solid lines are linear fits in the logarithmic space with a fixed slope of unity.}
 \label{fig6}
\end{figure}

 We plot the correlations between CH$_3$OCH$_3$, HCOOCH$_3$ and CH$_3$CH$_2$CN abundance ratios (relative to CH$_3$OH abundances) in Fig.~\ref{fig6}. The abundance ratios of all three species are clearly correlated with each other across 2-3 orders of magnitude, decreasing from hot corinos and EGO cloud cores to massive cloud cores (including OMC) to the GC clouds (including the cores in Sgr\,B2). The correlation between CH$_3$CH$_2$CN and CH$_3$OCH$_3$ is looser than between HCOOCH$_3$ and CH$_3$OCH$_3$. Especially, the figure shows that the organic abundance ratios form a continuous sequence over the different types of cloud cores and the EGO cloud cores are located roughly between the hot cores and hot corinos. Comparing these abundance ratios to the hot core stage in the chemical modeling of cloud warm-up by \citet{garr2013}, the observed organic abundance ratios do not agree to the model results. The typical hot core abundance ratios [HCOOCH$_3$/CH$_3$OH] and [CH$_3$OCH$_3$/CH$_3$OH] are both close to or smaller than $10^{-2}$ in the models, whilst most of the observed ratios in Fig.~\ref{fig6} are larger than this value. This discrepancy can not be reconciled unless most of the observed various cloud cores are all unbelievably crowed in the short onset phase of the hot core stage, as the EGO cloud cores are. Thus, it is more reasonable to conclude that the simple physical and chemical models of cloud warm-up still need to be improved to get the existing observations consistently interpreted.

The correlations in Fig.~\ref{fig6} roughly follow a slope of unity in logarithmic scale in both panels. Thus, we fit the correlations with straight lines with unity slope, which results in average abundance ratios between the three organic species among the various types of cloud cores as [CH$_3$OCH$_3$ : HCOOCH$_3$ : CH$_3$CH$_2$CN]$\approx$100:101:21. Here, the abundance of the first species has been normalized to the arbitrary number 100. The correlations and the abundance ratios over the diverse types of cloud cores may provide additional constraints to the chemical modeling of cloud core warm-up processes.

As mentioned earlier, the observed abundances can be very uncertain due to various sources of error such as different telescope beams, wavelengths, different partition functions (inclusion of vibrational states) and analysis methods (rotation diagram versus population diagram), telescope pointing errors and flux calibration errors, which makes comparison of chemical abundances between independent works difficult. However, by checking the different observation settings (telescope, beam, wavelength, and partition function) in Table~\ref{tab_clouds}, we do not see any apparent effects of these uncertain factors upon the abundance ratios or methanol abundances. Perhaps, the revealed trends of abundances and ratios are dominated by the intrinsic chemical variety among the different cloud cores.

\section{Summary}
\label{summary}
We perform a population diagram analysis to multiple observed lines of four organic species (CH$_3$OH, CH$_3$OCH$_3$, HCOOCH$_3$ and CH$_3$CH$_2$CN) of 29 EGOs that have at least reliable CH$_3$OH lines from our Paper\,I \citep{he2012}. The rotation temperatures, gas densities and cloud core masses derived from the methanol lines and the abundances of the four species allow us to discuss the physical and chemical properties of the EGO cloud cores. 

The EGO cloud cores show similar methanol $J_3-J_2$ rotation temperatures of about 44\,K on average. However, the gas kinetic temperatures could be higher due to non-thermal effects. The cores possibly have similar linear sizes of about 0.036\,pc. The gas densities and masses in the cores range from a few $10^5$ to a few $10^7$\,cm$^{-3}$ and a few tenth to a few tens solar masses, respectively.

The average (beam averaged) abundances of the four organic species are 
$3.20\times 10^{-9}$ for CH$_3$OH,
$1.36\times 10^{-9}$ for CH$_3$OCH$_3$, 
$7.55\times 10^{-9}$ for HCOOCH$_3$ and 
$3.94\times 10^{-10}$ for CH$_3$CH$_2$CN; 
the corresponding abundances in the small cores could be about 40 times higher than these ones. 
More elaborated average abundance ratios are obtained to be 
[CH$_3$OH : CH$_3$OCH$_3$ : HCOOCH$_3$ : CH$_3$CH$_2$CN] $\approx 200 : 61 : 172 : 15$. The abundances of all four species are correlated with each other, which supports the idea that they are produced through related chemical processes.
 
The organic abundances and abundance ratios suggest that these EGO cloud cores are neither in the early warm-up stage, nor in the later hot core stage, but possibly in the short onset phase of the hot core stage when methanol ice is quickly evaporating from grain surfaces at a gas temperature of about 100\,K and a chemical age of around $3.5\times10^4$yr. However, it is still puzzling why the infrared selected EGO cloud samples concentrate in so narrow ranges of gas temperature and age.

Comparisons between EGO cloud cores, hot corinos, massive hot cores, and GC clouds reveal that the abundances of CH$_3$OCH$_3$, HCOOCH$_3$ and CH$_3$CH$_2$CN relative to methanol are linearly correlated with each other over more than two orders of magnitude. These correlations imply abundance ratios of [CH$_3$OCH$_3$ : HCOOCH$_3$ : CH$_3$CH$_2$CN]$\approx$100:101:21. These general trends are independent of the physical conditions in the different types of clouds and thus should be related to the nature of the chemical networks. Both the abundance ratios and their correlations are yet to be explained by future chemical modeling. The chemical properties of the EGO cloud cores are between that of hot corinos and hot cores.

\section*{Acknowledgments}

Ge J.X., He J.H. and Chen X. thank the support of the Chinese National Science Foundation (Grant Nos. 11373067, 11173056, 11133008 and 11273043). 

\bibliography{ego-analysis} 
\bibliographystyle{mn2e} 

\appendix

\section[]{Example figures}
\label{examplefig}
\begin{figure*}
 \centering
  \includegraphics[angle=0,height=13cm,width=18cm]{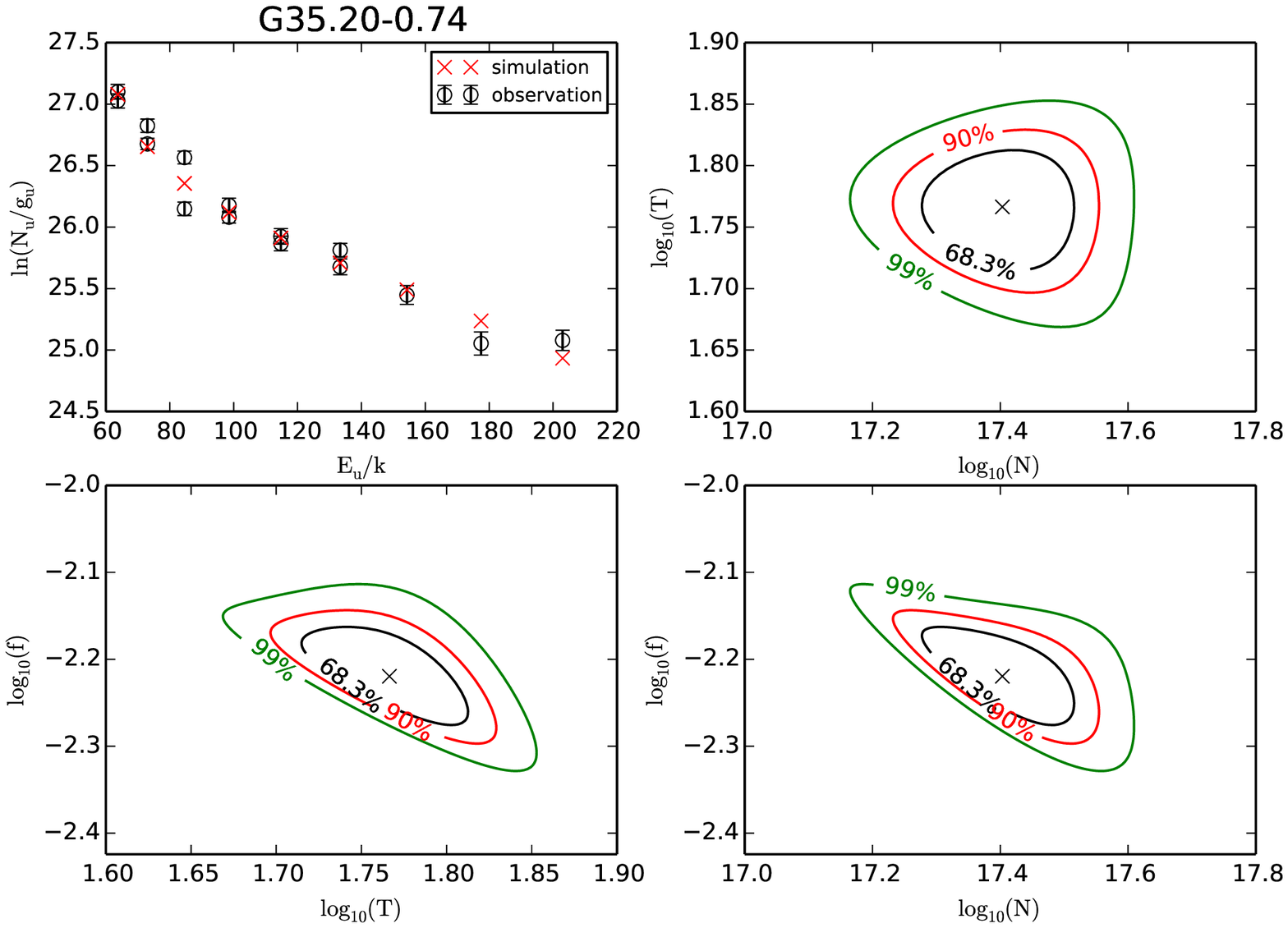}
 \includegraphics[angle=0,height=5cm,width=16cm]{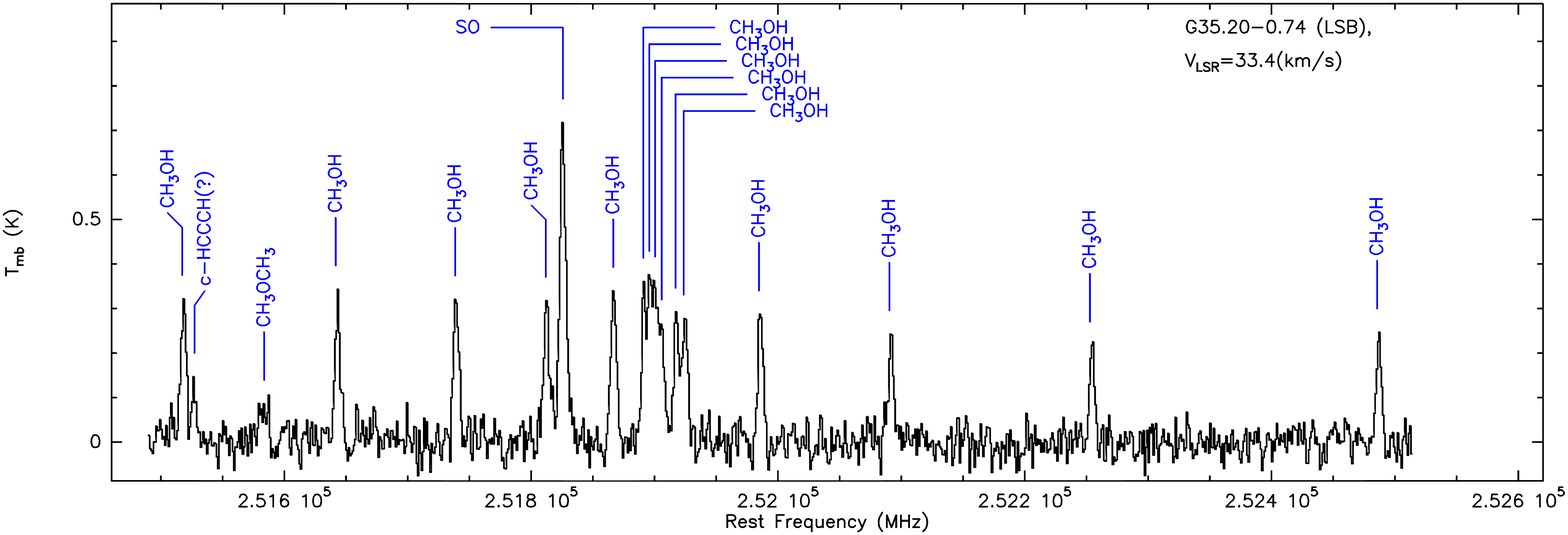}
 \caption{Example figure of the population diagram analysis, $\chi^2$ minimization and spectral plots of G35.20-0.74 from group-I in Table~\ref{tab_ch3oh} that contains both optically thick and thin CH$_3$OH lines.  The crosses in the contour figures correspond to the minimum value of $\chi$. A color plot is available in the electronic version.}
 \label{fig1}
\end{figure*}
\begin{figure*}
 \centering
  \includegraphics[angle=0,height=13cm,width=18cm]{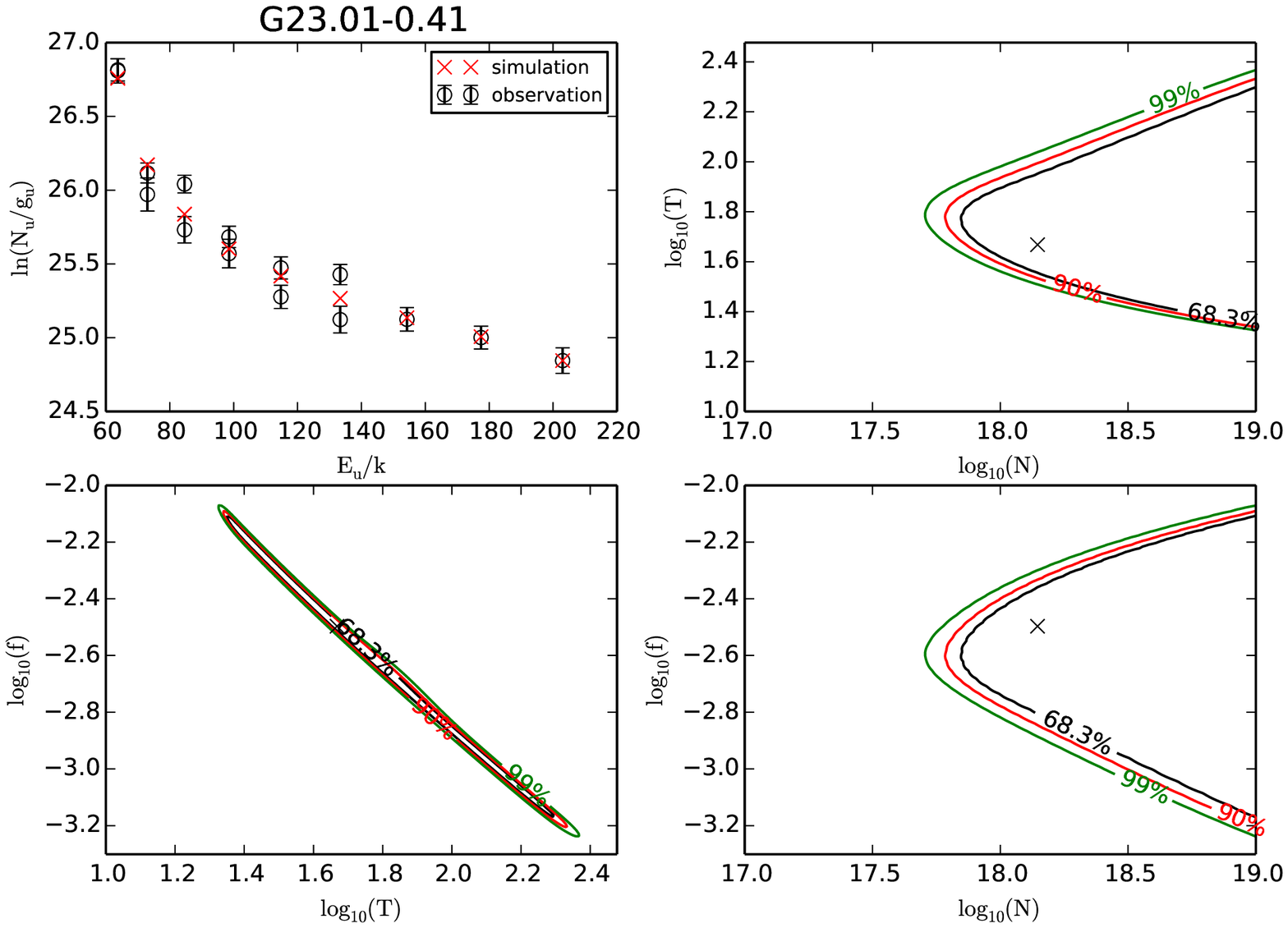}
  \includegraphics[angle=0,height=5cm,width=16cm]{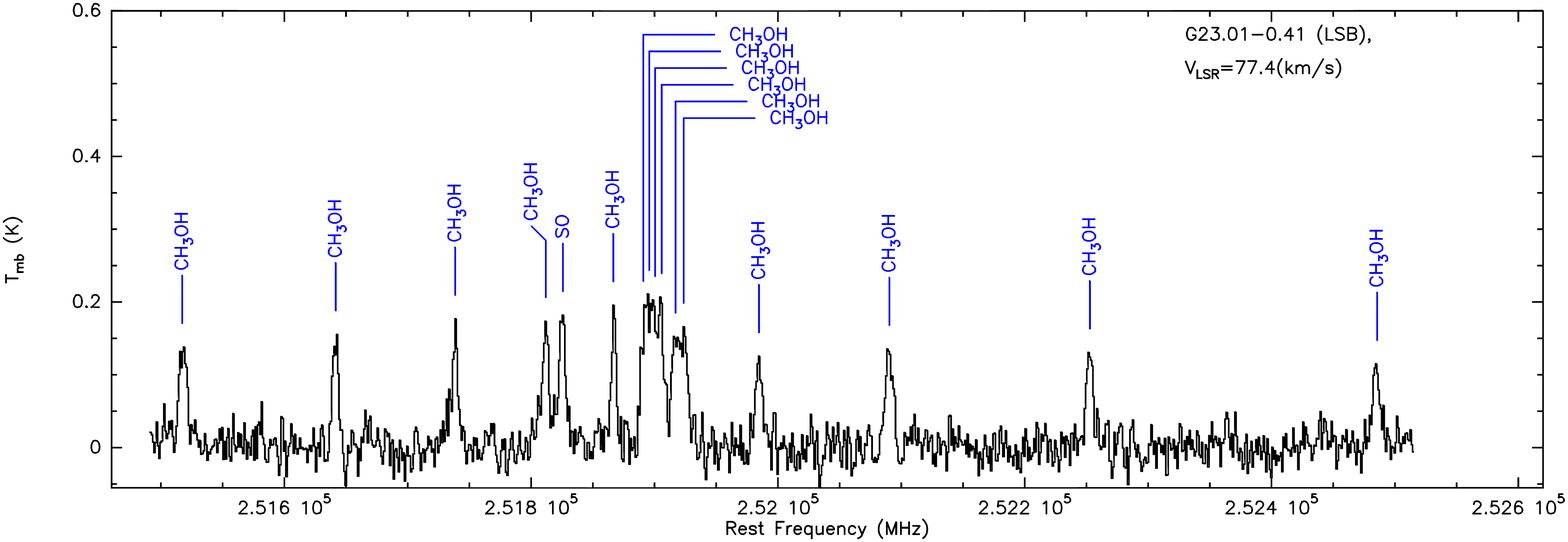}
 \caption{Example figure of the population diagram analysis, $\chi^2$ minimization and spectral plots of G23.01-0.41 from group-II in Table~\ref{tab_ch3oh} that only contains optically thick CH$_3$OH lines.  The crosses in the contour figures correspond to the minimum value of $\chi$. A color plot is available in the electronic version.}
 \label{fig2}
\end{figure*}

\begin{figure*}
 \centering
  \includegraphics[angle=0,height=13cm,width=18cm]{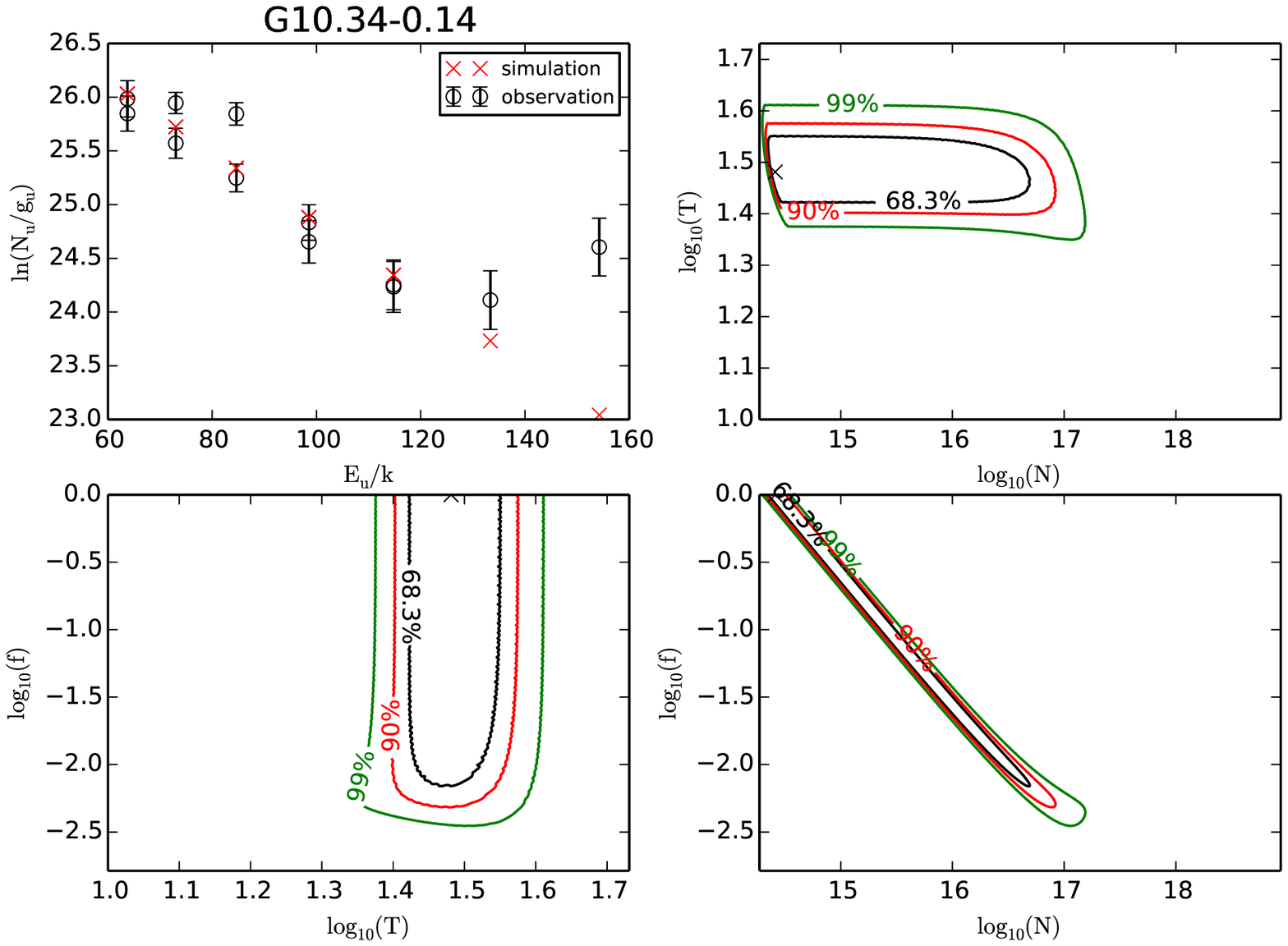}
 \includegraphics[angle=0,height=5cm,width=16cm]{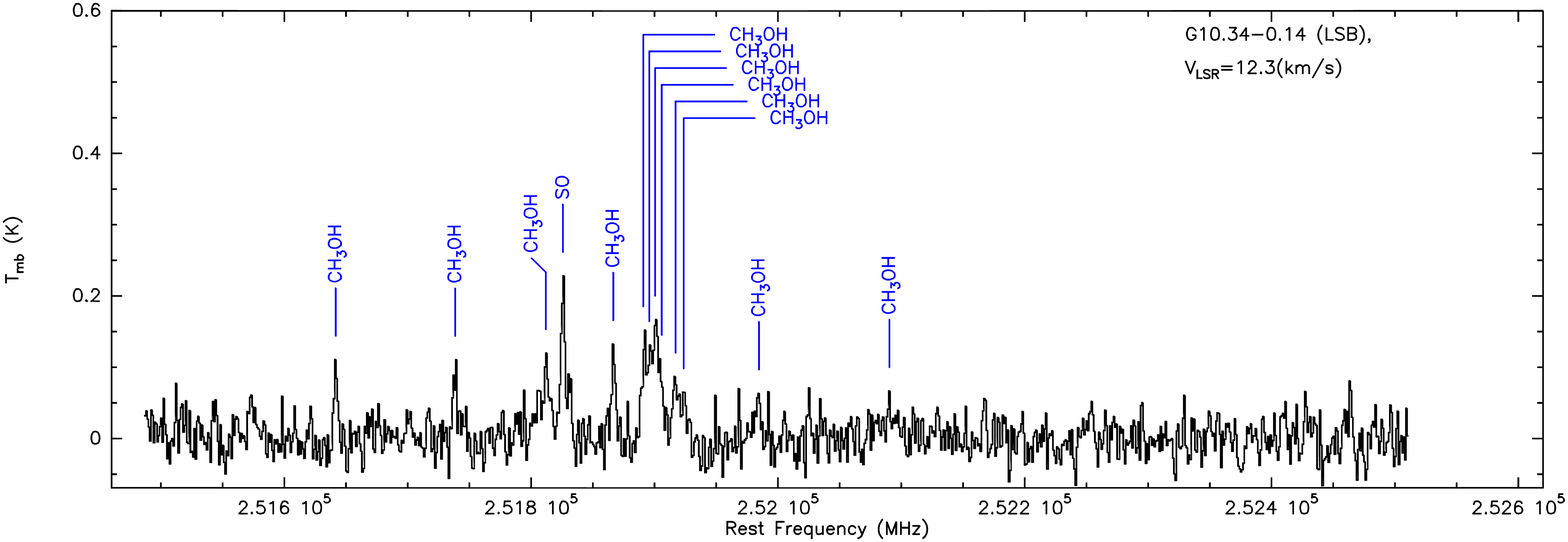}
 \caption{Example figure of the population diagram analysis, $\chi^2$ minimization and spectral plots of G10.34-0.14 from group-III in Table~\ref{tab_ch3oh} that only contains optically thin CH$_3$OH lines.  The crosses in the contour figures correspond to the minimum value of $\chi$. A color plot is available in the electronic version.}
 \label{fig3}
\end{figure*}

\bsp

\label{lastpage}

\end{document}